\documentclass[fleqn,usenatbib,useAMS]{mnras}

\usepackage[T1]{fontenc}
\usepackage{ae,aecompl}

\usepackage{bm}
\usepackage{graphicx}	
\usepackage{amsmath}	
\usepackage{amssymb}	
\usepackage{gensymb}

\usepackage{txfonts}

\title[Evolution of $\alpha$ Centauri~B's protoplanetary disc]{Evolution of $\alpha$ Centauri B's protoplanetary disc}

\author[Martin, Lissauer \& Quarles]{Rebecca G. Martin$^{1}$, Jack J. Lissauer$^2$
and Billy Quarles$^3$ 
\\$^{1}$Department of Physics and Astronomy, University of Nevada, Las Vegas, 4505 South Maryland Parkway, Las Vegas, NV 89154, USA \\$^2$Space Science \& Astrobiology Division, MS 245-3, NASA Ames Research Center, Moffett Field, CA 94035, USA \\$^3$Center for Relativistic Astrophysics, School of Physics, 
Georgia Institute of Technology, Atlanta, GA 30332 USA}

\date{Accepted XXX. Received YYY; in original form ZZZ}

\pubyear{2020}

\begin{document}
\label{firstpage}
\pagerange{\pageref{firstpage}--\pageref{lastpage}} 
\maketitle

\begin{abstract}
With hydrodynamical simulations we examine the evolution of a protoplanetary disc around $\alpha$ Centauri~B including the effect of the eccentric orbit binary companion $\alpha$ Centauri~A. The initially circular orbit disc undergoes two types of eccentricity growth. First, the eccentricity oscillates on the orbital period of the binary, $P_{\rm orb}$, due to the eccentricity of the binary orbit. Secondly, for a sufficiently small disc aspect ratio, the disc undergoes global forced eccentricity oscillations on a time-scale of around $20\,P_{\rm orb}$. These oscillations damp out through viscous dissipation leaving a quasi-steady eccentricity profile for the disc that oscillates only on the binary orbital period. 
The time-averaged global eccentricity is in the range 0.05-0.1, with no precession in the steady state.  The periastrons of the gas particles are aligned to one another. The higher the disc viscosity, the higher the disc eccentricity. With $N$-body simulations we examine the evolution of a disc of planetesimals that forms  with the orbital properties of the quasi-steady protoplanetary disc.  We find that the average  magnitude of the eccentricity of particles increases  and their periastrons become misaligned to each other once they decouple from  the gas disc. The low planetesimal collision velocity required for planet formation suggests that for planet formation to have occurred  in a disc of planetesimals formed from a protoplanetary disc around $\alpha$ Centauri B, said disc's viscosity must be have been small and planet formation must have occurred at orbital radii smaller than about $2.5\,\rm au$.  Planet formation may be easier with the presence of gas.
\end{abstract}

 
\section{Introduction}
\label{intro}

The triple star system $\alpha$ Centauri is the closest star system to the solar system.  $\alpha$ Centauri~A, with mass $M_A=1.133\,\rm M_\odot$, and $\alpha$ Centauri~B, with mass $M_B=0.973\,\rm M_\odot$, are a relatively close binary with orbital period of $79.91\,\rm yr$ and eccentricity $e_{\rm b}=0.51$ \citep{Pourbaix2016}.  The third component, $\alpha$ Centauri~C (also known as Proxima Centauri), is a small and faint red dwarf of mass $M_C=0.15\,\rm M_\odot$ that is loosely-bound to $\alpha$ Centauri AB, with an estimated orbital semimajor axis $\sim$~9,000 AU from the center of mass \citep{Kervella2017}.

Numerous groups have searched for planets in the $\alpha$ Centauri ABC
system during the past two decades \citep[e.g.][]{Endl2001,Milli2013,Bergmann2015,Endl2015,Clery2018}.   Radial velocity studies have revealed a 
 planet with mass $M_p\sin i \approx 1.27\,\rm M_\oplus$ and orbital period of 11.2 days orbiting Proxima Centauri \citep{AngladaEscude2016}.  This planet, $\alpha$ Centauri~C~b, orbits with a semi--major axis of $0.049\,\rm au$, placing it within the habitable
zone of its faint stellar host. A low-mass planet in a short period orbit about $\alpha$ Centauri~B was announced by a radial velocity group in 2012, but later shown to be an artifact
of the analysis \citep{Hatzes2013,Rajpaul2016}. Very tentative evidence for a possible transiting planet around $\alpha$ Centauri~B has also been presented \citep{Demory2015}. Observations so far have failed to find any evidence
for giant planets. Radial velocity measurements rule out planets with
a mass greater than $8.4\,\rm M_\oplus$ orbiting interior to the outer boundary of  $\alpha$ Centauri~B's habitable zone and planets with
a mass exceeding $53\,\rm M_\oplus$ orbiting interior to the outer boundary of  $\alpha$ Centauri~A's habitable zone \citep{Zhao2018}. \cite{Trigilio2018} found no evidence for magnetic star--planet interaction around either $\alpha$ Centauri~A or B.  Observations in the near future with the upcoming James Webb Space Telescope (JWST) and ground based facilities will constrain the parameter space further \citep{Beichman2020}.

Proxima Centauri is so far away from the binary $\alpha$~Centauri~AB that it would not significantly affect disc evolution or planet formation around either component \citep[e.g.][]{Worth2016}. However, A and B will significantly affect the planet formation process around each other. \cite{Quintana2002} used N--body simulations to simulate the late stages of terrestrial planet formation around $\alpha$ Centauri A and B. They started with a gas-free disc of Mars--Moon sized bodies and found that planet formation occurs provided that the disc is not so highly inclined to the binary orbital plane that it becomes unstable to Kozai--Lidov oscillations \citep{Kozai1962,Lidov1962}. The planetary accretion time-scales that they found are shorter than those around a single sun--like star lacking giant planets, but similar to those computed for the terrestrial planets in our own solar system with Jupiter and Saturn included \citep{Quintana2014}.  However, all of Quintana et al.'s simulations assumed discs of Mars-sized and Moon-sized bodies similar to those used for modeling terrestrial planets in our solar system with planetary embryos beginning on nearly circular orbits \citep[see also simulations by][]{Barbieri2002,Guedes2008,Thebault2008,Thebault2009,Xie2010}. \cite{Quarles2016,Quarles2018b} showed that Earth mass planets, and even systems of multiple Earth mass planets, could survive on orbits within the habitable zones of both $\alpha$ Centauri~A and B on gigayear time-scales   \citep[see also][]{Benest1988,Wiegert1997,Popova2012,AndradeInes2014}.

Formation of planets from the collisions of planetestimals around one component of a close and eccentric binary may be difficult \citep{Rafikov2013,Rafikov2015,Rafikov2015b,Silsbee2015}.
The planetesimals are subject to eccentricity excitation that leads to high relative velocities between the planetesimals and thus destructive collisions \citep{Paardekooper2008,Kley2008,Paardekooper2010,Marzari2012}. The gas disc may provide a drag that is able to align the orbital eccentricities of the planetesimals thus reducing the relative speeds of collisions \citep{Marzari2000}. The effect of a gas disc 
may overcome these issues if the mass of the gas disc is large and the eccentricity low, $e\lesssim 0.01$ \citep{Rafikov2015}.

In this work we are interested in how a protoplanetary disc
around one of the close binary components is affected by the presence
of the companion and how this affects the initial conditions for planet
formation. The binary companion drives eccentricity growth in the disc \citep[e.g.][]{Lubow1991,Lubow1991b,Kley2008,KPO2008}. The companion drives strong spiral arms at periastron that propagate inwards through the disc \citep[e.g.][]{Muller2012,Picogna2013}.

We present herein the results of three-dimensional hydrodynamical simulations to
investigate the evolution a protoplanetary disc around $\alpha$ Centauri B and
with N-body simulations we explore the implications for planet formation therein.  We first examine individual test particle orbits
around $\alpha$ Centauri~B in
Section~\ref{particleorbits}.  An initially circular orbit displays forced
eccentricity oscillations due to the large eccentricity of the
binary. In Section~\ref{hydro} we use hydrodynamic disc simulations to
show that the disc around $\alpha$ Centauri~B is able to display global forced
eccentricity oscillations but these damp in time because of the
viscosity of the disc. The disc eventually reaches a quasi-steady
state with an eccentricity distribution over the radial extent of the
disc. In Section~\ref{discussion} we discuss the implications of our results for planet formation around $\alpha$~Centauri~A and $\alpha$~Centauri~B. We use $N$-body simulations to explore the evolution of planetesimals that form from the protoplanetary disc. We draw our conclusions in Section~\ref{conc}.

\section{Test Particle Orbits}
\label{particleorbits}

\begin{figure*}
\begin{center}
\includegraphics[width=0.48\textwidth]{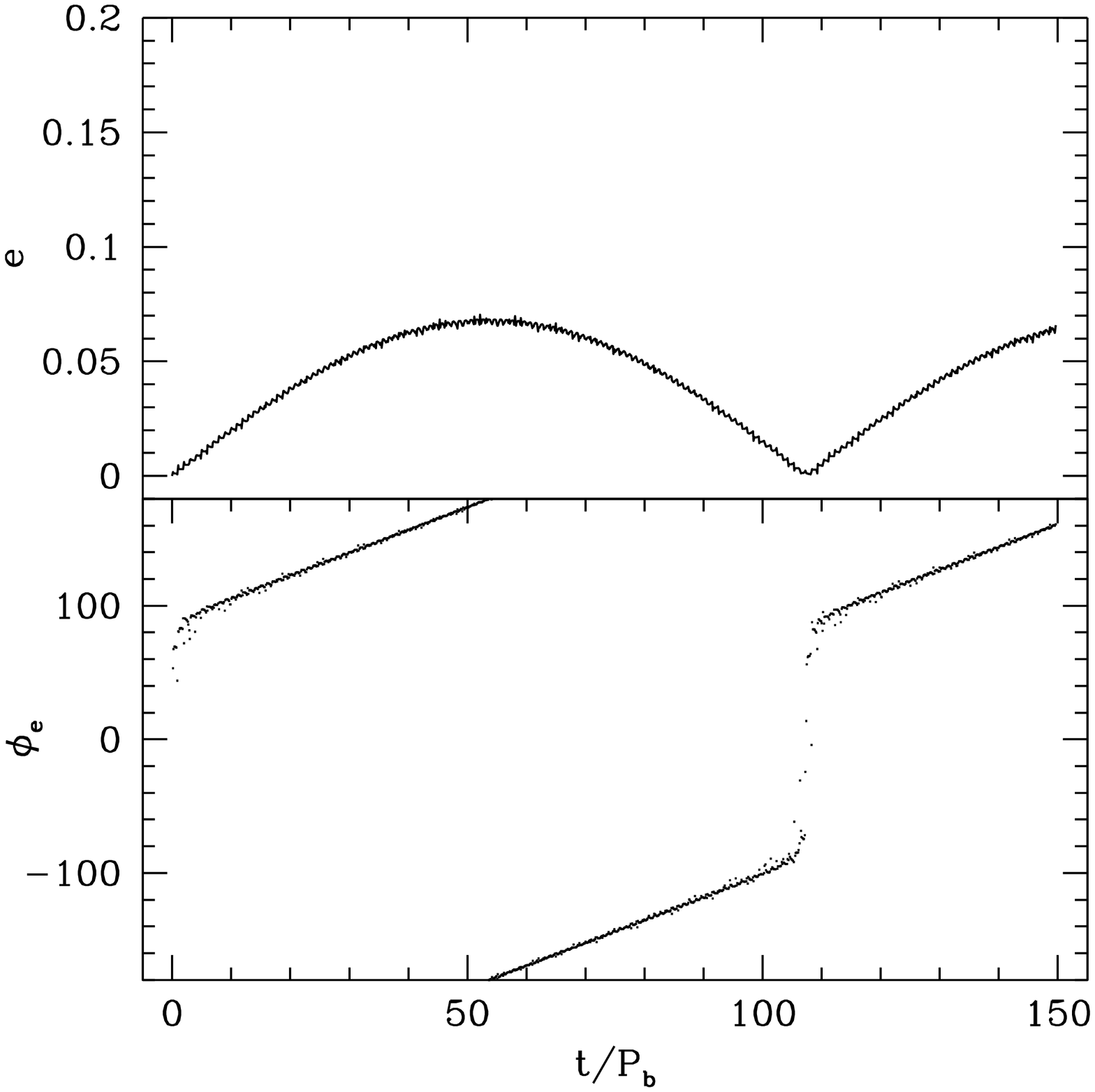}
\includegraphics[width=0.48\textwidth]{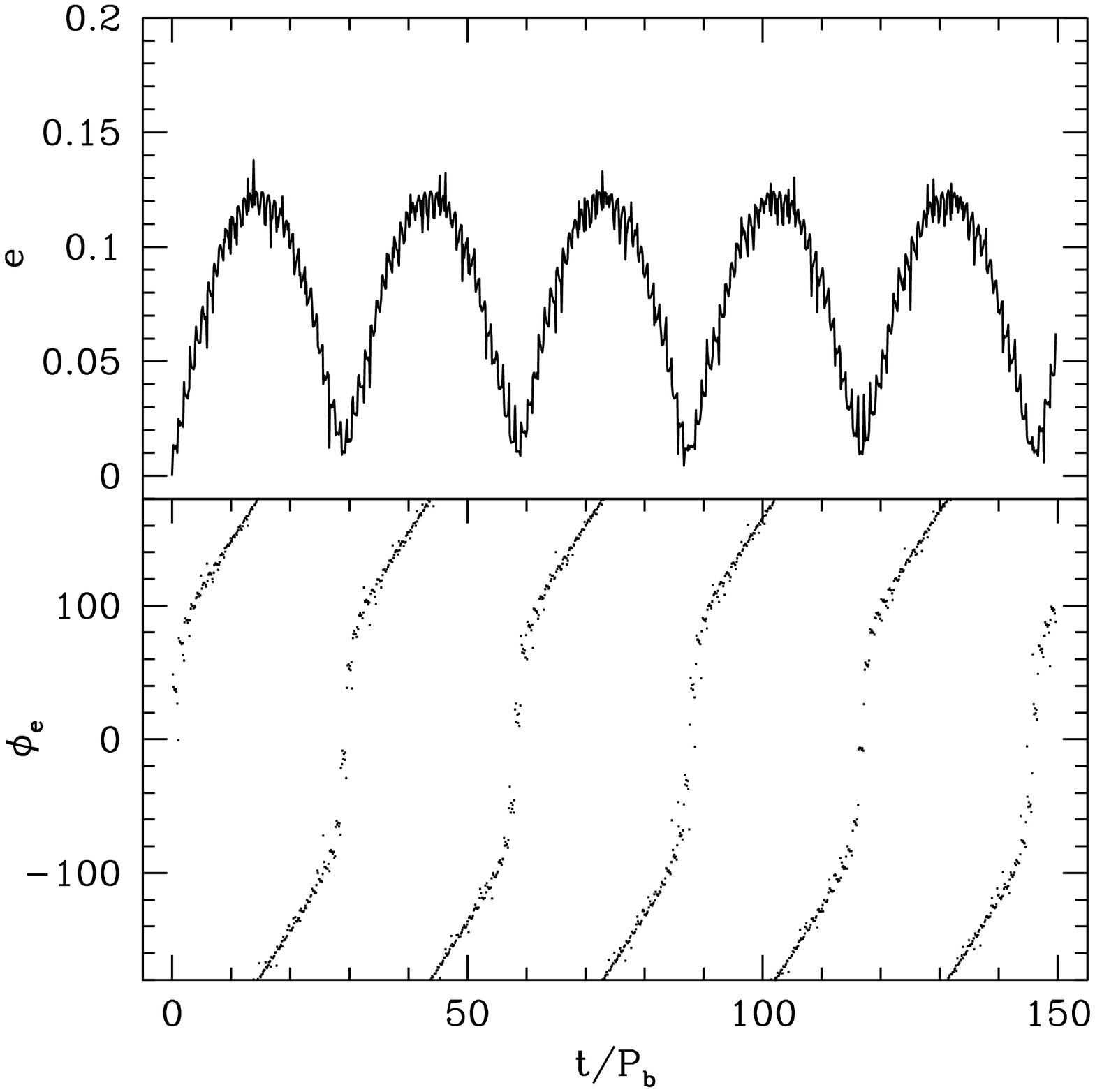}
\includegraphics[width=0.48\textwidth]{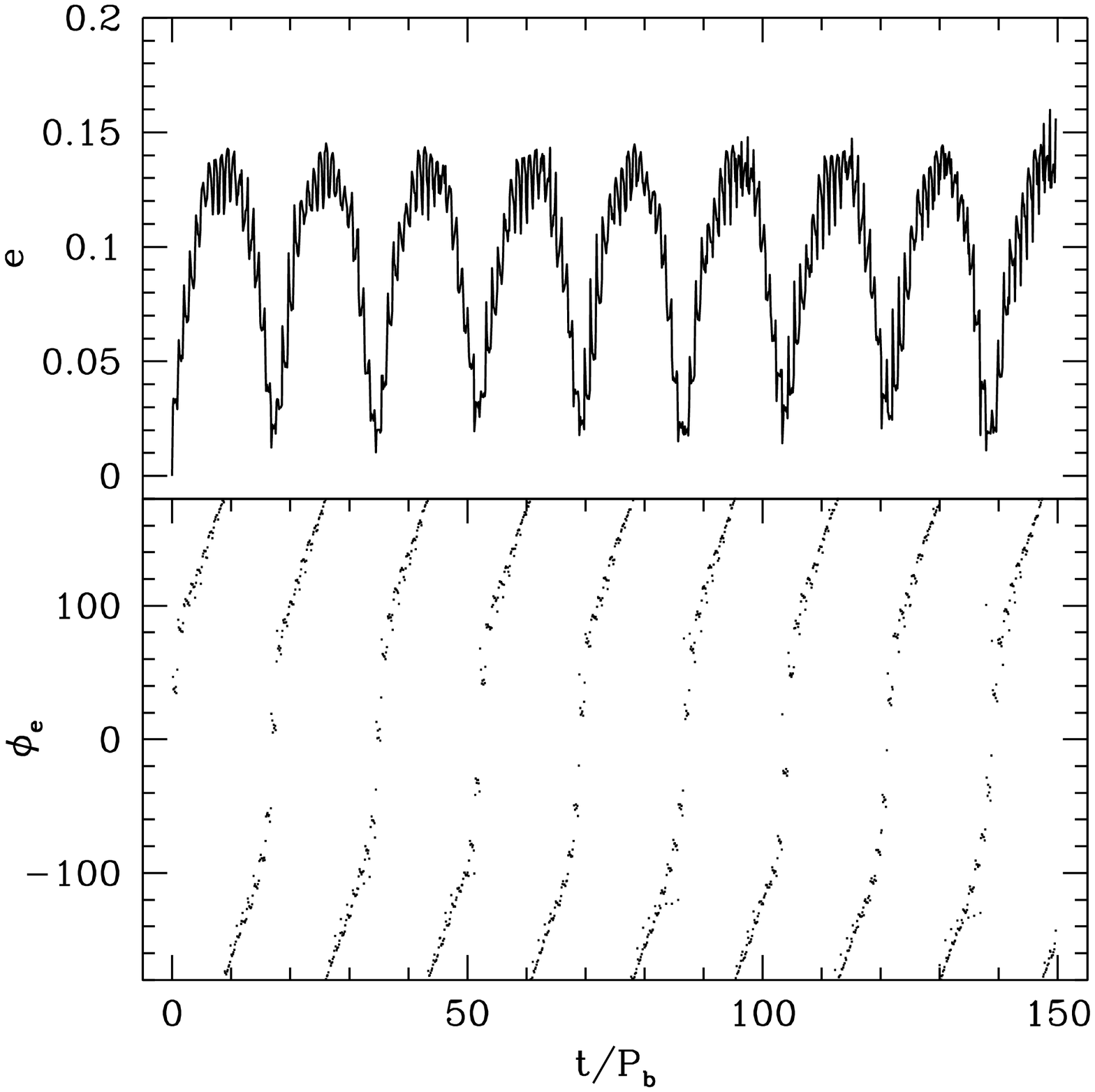}
\includegraphics[width=0.48\textwidth]{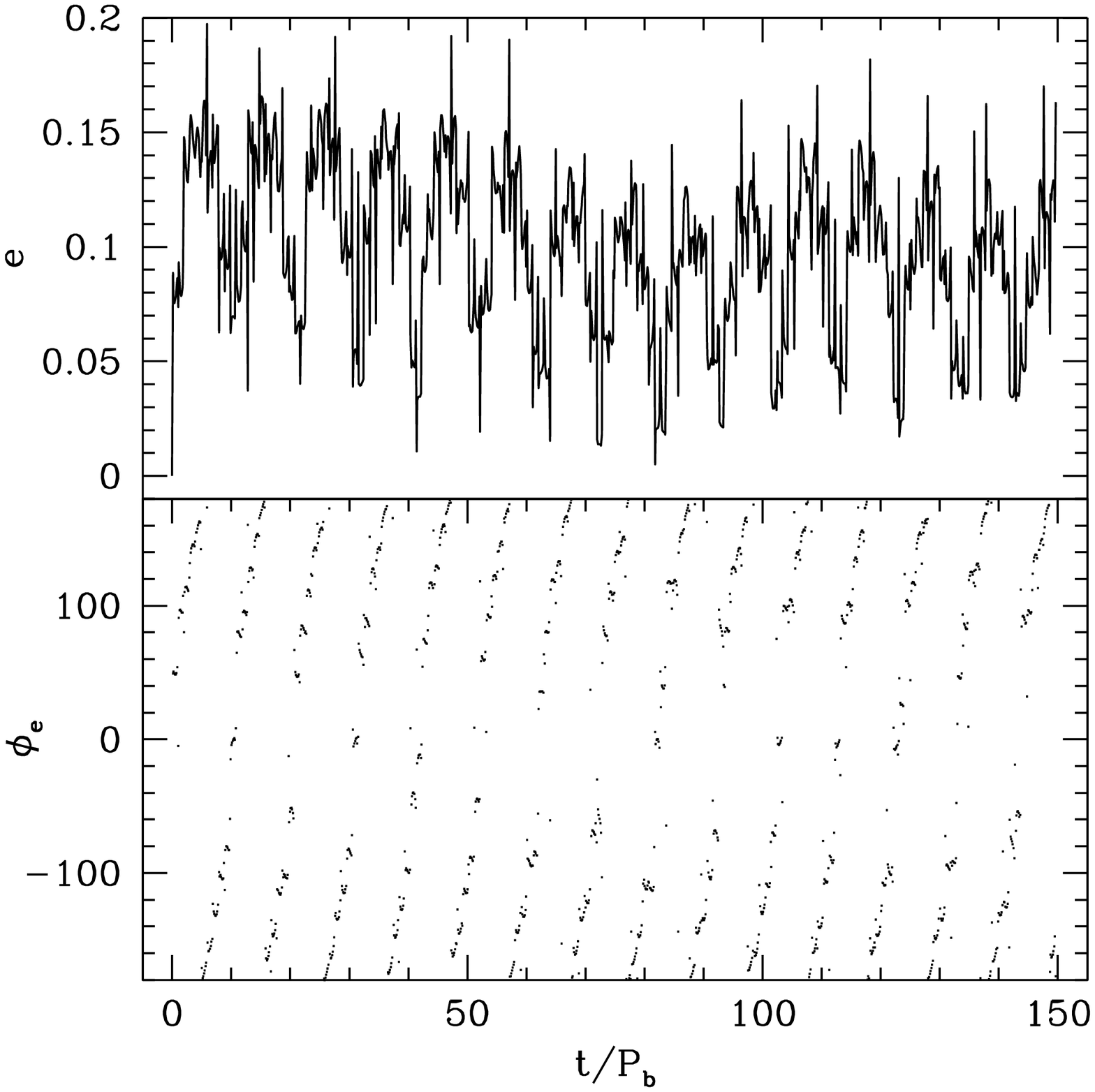}
\end{center}
\caption{The eccentricity evolution of an initially circular test
  particle around $\alpha$ Centauri~B for four different initial radii
  $R=1\,\rm au$ (top left), $R=2\,\rm au$ (top right), $R=2.5\,\rm au$
  (bottom left) and $R=3\,\rm au$ (bottom right). The upper panels show the magnitude of the eccentricity while the lower panels show the eccentricity apsidal angle. 
  }
\label{particle}
\end{figure*}

We first consider test particle orbits around $\alpha$ Centauri~B. The test
particles begin in a circular Keplerian orbit with zero eccentricity. The test particle orbit is coplanar to the binary orbit and remains so during the simulation.   The eccentricity of the particle is defined by a vector $\bm{e}=(e_x,e_y,e_z)$. The magnitude of the eccentricity is $e=|\bm{e}|$ and the phase angle (apsidal angle) is defined as
\begin{equation}
\phi_e=\tan^{-1}\left(\frac{e_y}{e_x}\right).
\label{phie}
\end{equation}
Fig.~\ref{particle} shows the  evolution of the magnitude of the particle orbital eccentricity and the eccentricity apsidal angle for particles of
varying initial separation from the star. 

At small orbital radius, close to $\alpha$ Centauri~B, the test particles display periodic forced eccentricity oscillations. The magnitude of the eccentricity growth increases with distance from the central star. The time-scale for the oscillations decreases farther from the central star.  The oscillations become more chaotic for particles that begin with a semi--major axis that is farther from the host star, or closer to the perturbing star.   The eccentricity vector for the particle precesses in time. The closer the particle orbit is to the perturbing binary companion, the faster the precession. Because the orbit begins with zero eccentricity, there are jumps in the apsidal angle of $\pi$ every time the magnitude of the eccentricity passes through zero.

\begin{figure}
\begin{center}
\includegraphics[width=0.45\textwidth]{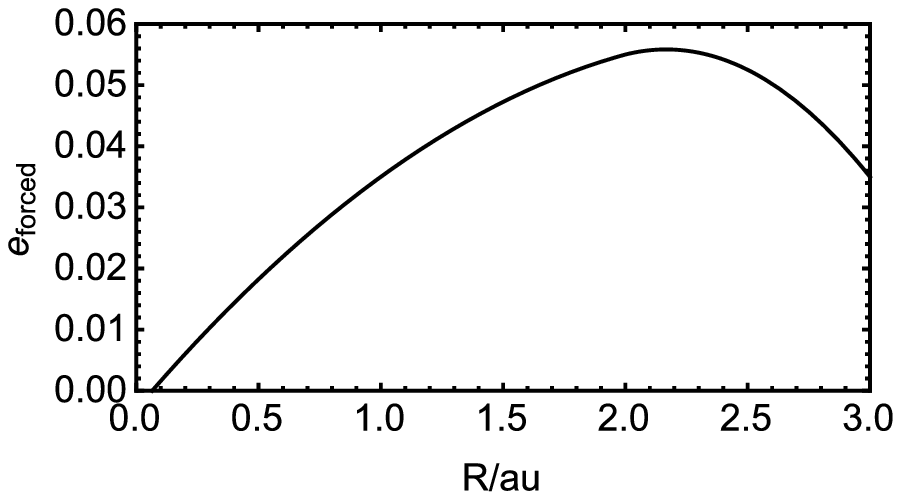}
\end{center}
\caption{The forced eccentricity of a test particle as a function of its separation from the host star. The fit is provided in \protect\cite{Quarles2018}.}
\label{particle2}
\end{figure}

\cite{AndradeInes2017} and \cite{Quarles2018} presented extensive studies of the evolution of small planets on prograde, planar orbits about $\alpha$ Centauri B.  They found that perturbations from  $\alpha$ Centauri~A excite forced eccentricities that increase with semi--major axis up to a value of 0.06 near $a = 2.4\,$au, outwards of which effects of resonant perturbations complicate the dynamics \citep[see also][]{AndradeInes2016}. Figure~\ref{particle2} shows the forced eccentricity, $e_{\rm forced}$, of a test particle with the numerical fit provided in Table 3 of \cite{Quarles2018b}. The maximum eccentricity obtained by an initially circular orbit test particle is about $2e_{\rm forced}$, as seen by comparing with Fig.~\ref{particle} for test particles that begin with semi--major axis $R\lesssim 2\,\rm au$.

In the next Section we consider the response of a hydrodynamical disc (including pressure and viscous internal forces) that begins with particles in circular orbits. Without the effects of pressure and viscosity, the particles in the disc would undergo the same eccentricity oscillations as the test particle orbits described in this Section.

\section{Hydrodynamic disc simulations}
\label{hydro}

\begin{table*}
\caption{Parameters of the disc and binary in the hydrodynamical simulations. The first column is the simulation name. The second column is the binary eccentricity. The third column is the disc inner radius (which is equal to the accretion radius of the central star, $R_{\rm acc}$). The fourth column is the initial disc outer radius. The fifth column is the viscosity parameter. The sixth column is the disc aspect ratio at $R=0.5\,\rm au$. The seventh column is the initial number of SPH particles in the simulation. The eighth column is the artificial viscosity. The ninth column is the initial average smoothing length. The tenth column is the density weighted global disc eccentricity averaged over time $t=14\,P_{\rm orb}$ to $t=15\,P_{\rm orb}$. The eleventh column is the density weighted disc eccentricity in $R<2.5\,\rm au$ averaged over time $t=14\,P_{\rm orb}$ to $t=15\,P_{\rm orb}$. Note that these two averages for run7 are calculated over $t=49\,P_{\rm orb}$ to $t=50\,P_{\rm orb}$. The numbers in bold highlight the physical parameter that has been changed compared to the fiducial simulation, run1b.} \centering
\begin{tabular}{lcccccccccc} 
\hline 
Simulation & $e_{\rm b}$ & $R_{\rm in}\rm /au$ &  $R_{\rm out}\rm /au$ & $\alpha$ & $H/R (0.5{\,\rm au})$ & N & $\alpha_{\rm AV}$ &  $\left<h\right>/H$ & $\overline{\left<e\right>}$ & $\overline{\left<e\right>}_{\rm inner}$\\
\hline
\hline
run1&0.51&  0.2 & 3 & 0.01 & 0.1 & {\bf 1} $\times$ {\bf 10}$^5$ & 0.19  & 0.53 & 0.083 & 0.084\\
run1b &0.51&  0.2 & 3 & 0.01 & 0.1 & $3\times 10^5$ & 0.27 & 0.37 & 0.065 & 0.065 \\
run1c&0.51&  0.2 & 3 & 0.01 & 0.1 & {\bf 1} $\times$ {\bf 10}$^6$ & 0.40 & 0.25 & 0.059 &  0.057\\
\hline
single1 & {\bf -} & 0.2 & 3 & 0.01 & 0.1 & $3\times 10^5$&0.28& 0.37 & 0.018 & 0.016 \\
circular1 & {\bf 0} & 0.2 & 3 & 0.01 & 0.1 & $3\times 10^5$& 0.28 & 0.37 & 0.052 & 0.019\\
run2 &0.51&  {\bf 0.1} & 3 & 0.01 & 0.1 & $3\times 10^5$ &0.26 &0.38 & 0.063 & 0.062\\
run3 &0.51&  {\bf 0.5} & 3 & 0.01 & 0.1 & $3\times 10^5$ &0.29 & 0.34 & 0.071 & 0.072 \\
run4 &0.51 & 0.2 & {\bf 2} & 0.01 & 0.1 & $3\times 10^5$ & 0.30& 0.34 & 0.065 & 0.063 \\
run5 &0.51 & 0.2 & {\bf 4} & 0.01 & 0.1 & $3\times 10^5$ & 0.25 & 0.39 & 0.062 & 0.061 \\
run6 &0.51 &  0.2 & 3 & 0.01 &{\bf 0.075}& $3\times 10^5$ & 0.22 & 0.45 & 0.071 & 0.081 \\
run7 &0.51 & 0.2 & 3 & 0.01 & {\bf 0.05} & $3\times 10^5$ &0.17 &0.59 & 0.050 & 0.052 \\
run8 & 0.51 & 0.2& 3  & {\bf 0.05}&0.1  & $3\times 10^5$ &1.35 & 0.37 & 0.12 & 0.11 \\
\hline
\end{tabular}
\label{tab}
\end{table*}

\begin{figure*}
\begin{center}
\includegraphics[width=0.325\textwidth]{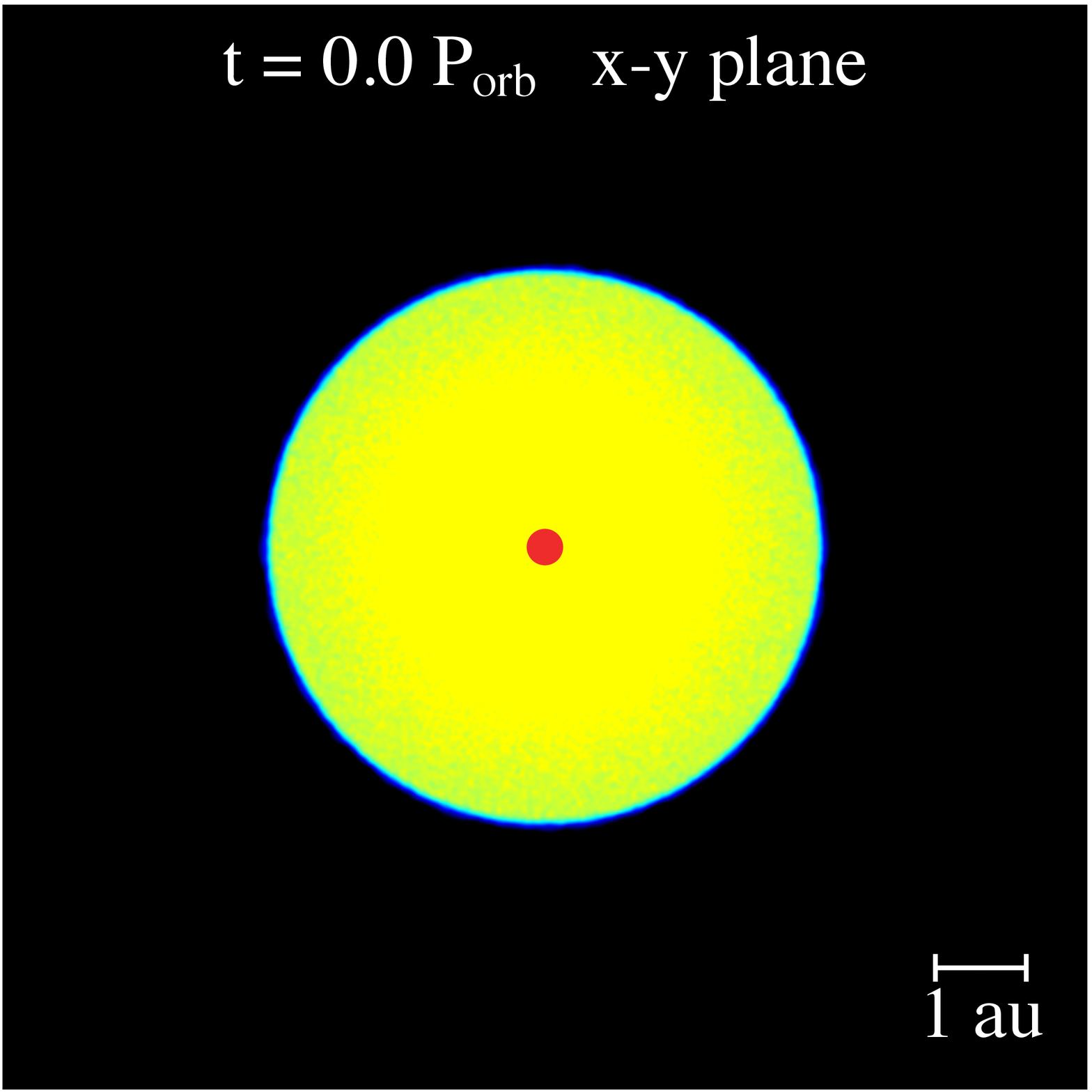}
\includegraphics[width=0.325\textwidth]{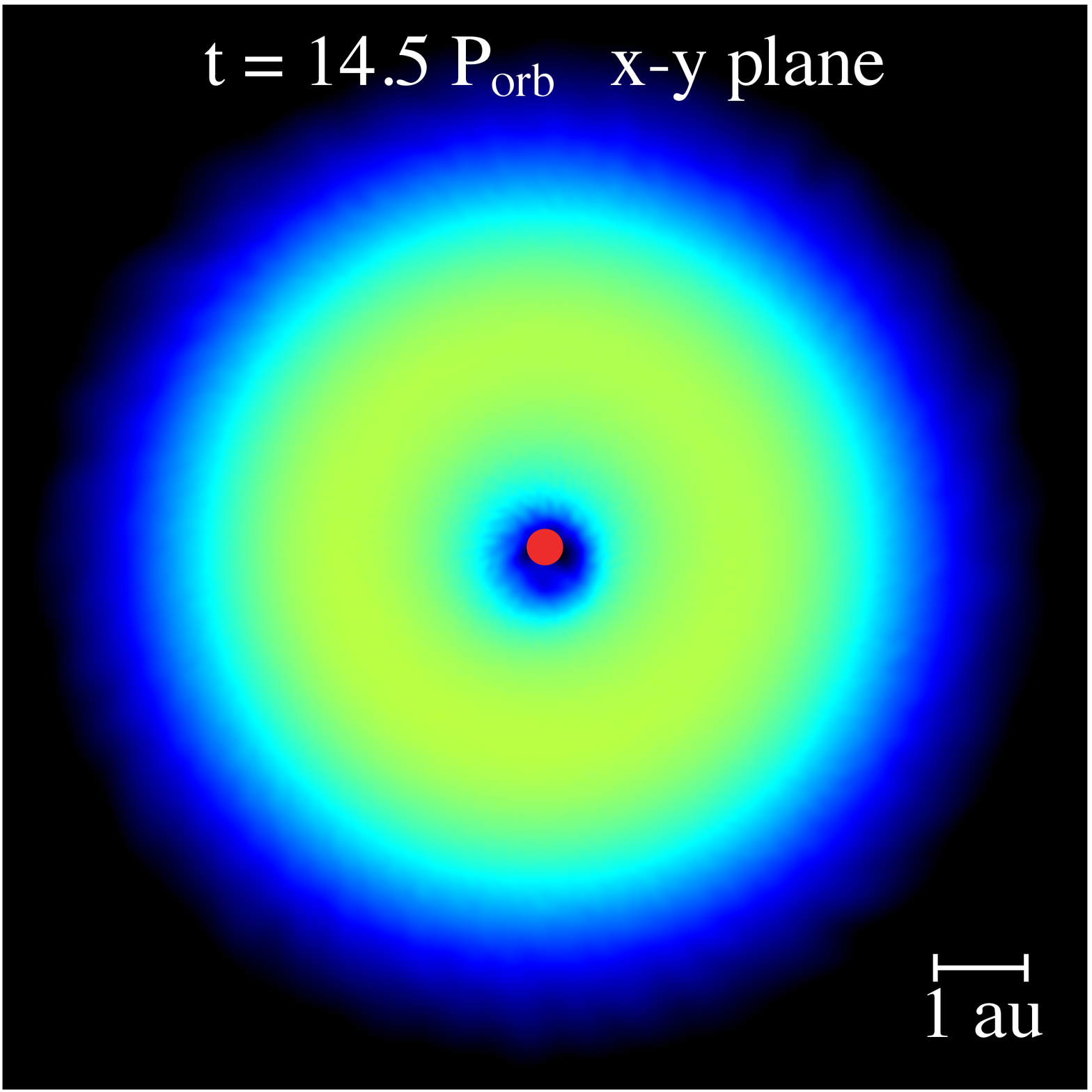}
\includegraphics[width=0.325\textwidth]{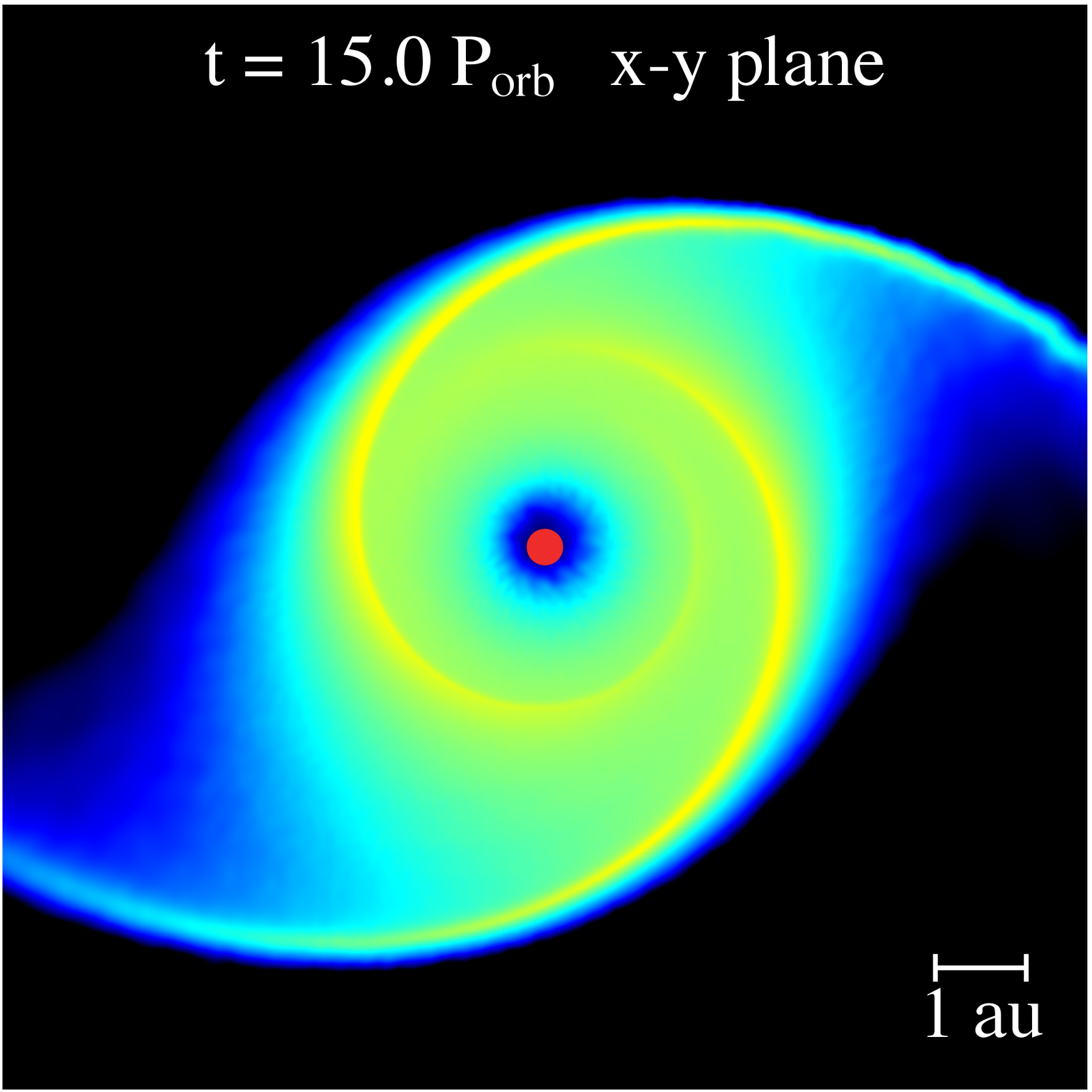}
\includegraphics[width=0.325\textwidth]{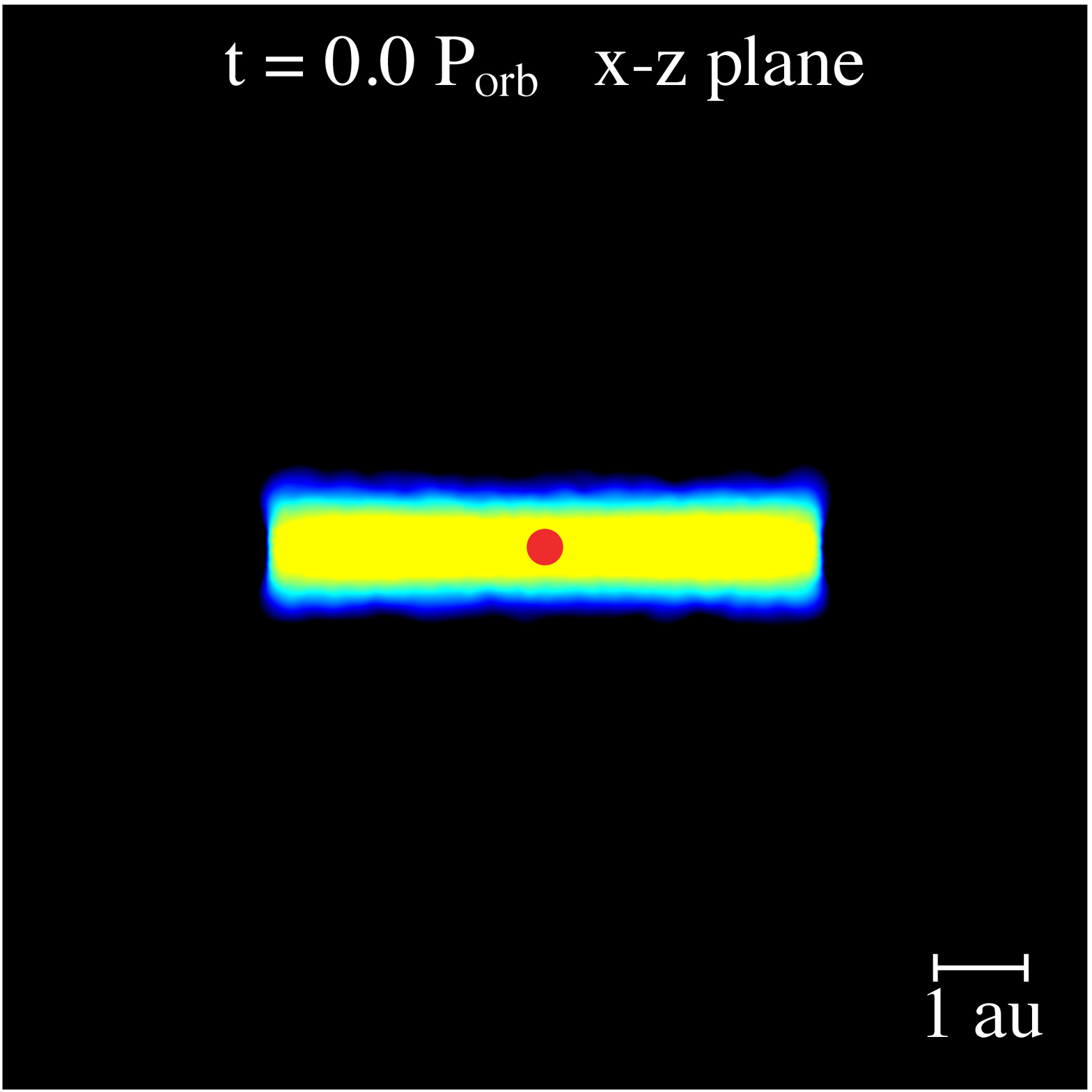}
\includegraphics[width=0.325\textwidth]{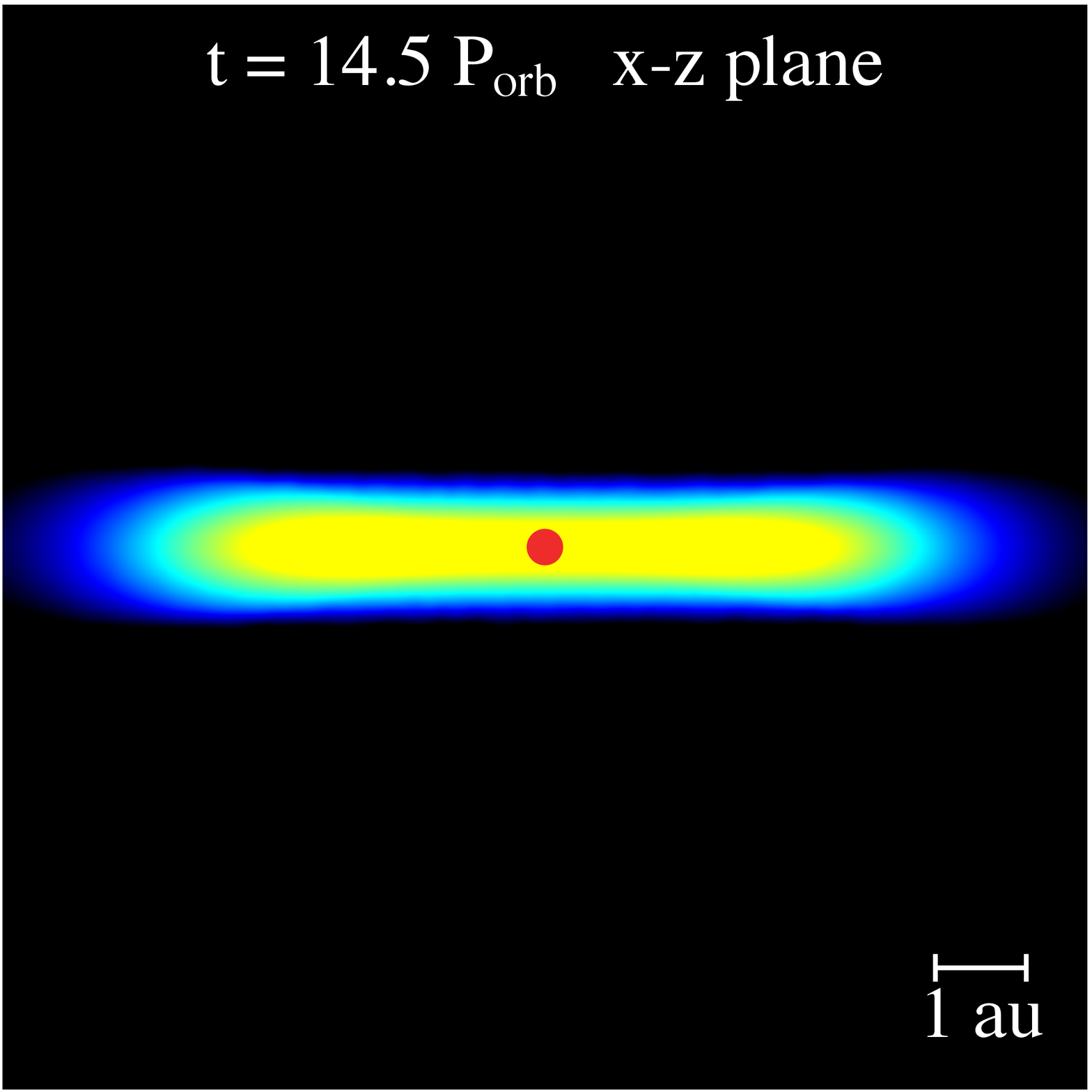}
\includegraphics[width=0.325\textwidth]{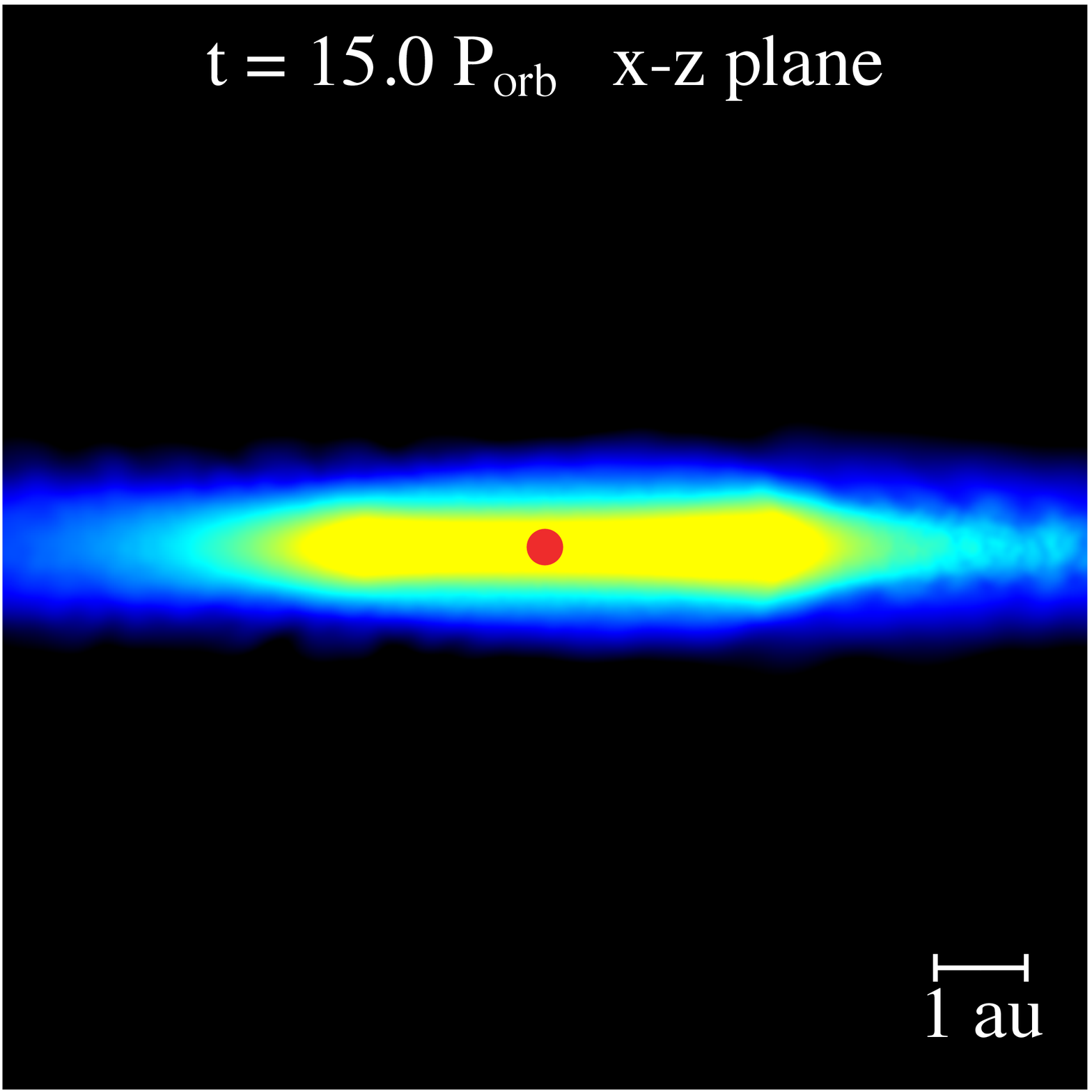}
\end{center}
\caption{Disc column density evolution in a frame that is corotating
  with the binary for the fiducial disc model (run1c). The colour of the gas denotes the column density, with yellow being about two orders of magnitude larger than blue.  The small red circle at the origin shows the
  disc's host star, $\alpha$~Centauri~B, with the size scaled to its
  accretion radius. The companion star is along the positive $x$--axis,
  but is not shown. The plots show times of $t=0$ (left),
  $t=14.5\,P_{\rm orb}$ (middle) and $t=15.0\,P_{\rm orb}$ (right).  The upper panels show the $x-y$ plane in which the binary orbits and the lower panels show the $x-z$ plane.
  }
\label{disc}
\end{figure*}

We investigate the evolution of a circumstellar gas disc around $\alpha$ Centauri~B. We use the smoothed particle hydrodynamics (SPH) code {\sc phantom} \citep{PF2010,LP2010,Price2012a,Nixon2012,Nixonetal2013,Price2018}. The
simulation parameters are summarised in Table~\ref{tab}. The masses
are $M_A=1.133\,\rm M_\odot$, $M_B=0.973\,\rm M_\odot$ and the mass of
the disc is $0.001\,\rm M_\odot$. Since the disc mass is small and it has little effect on the evolution, we do not include self--gravity into our simulations. Note, however, that the effect of self--gravity may be to decrease the eccentricity growth of the disc \citep{Marzari2009}. For the Toomre $Q$ parameter to be less than 2, the disc mass must be $M_{\rm d}\gtrsim 0.04 M_\odot$. This is a high disc mass since the radial extent of the disc is small due to tidal truncation by the binary.  The eccentricity of the binary is $e_{\rm b}=0.51$ and the
semi-major axis is $a_{\rm b}=23.75\,\rm au$.  The binary begins initially at periastron separation. Each star has a defined accretion radius, $R_{\rm acc}$. If material
passes within this radius it is accreted on to the sink particle (i.e., the star). The
surface density is initially distributed as $\Sigma \propto R^{-3/2}$,
and the disc initially extends from $R_{\rm in}=R_{\rm acc}$ out to
$R_{\rm out}$. However, the disc may spread outwards farther than its
initial outer radius due to viscous evolution.  The particles are distributed in the vertical direction with a Gaussian distribution with thickness $H=c_{\rm s}/\Omega$, where $c_{\rm s}$ is the sound speed and $\Omega$ is the Keplerian angular velocity.

The tidal truncation radius for a cold disc around one component of a circular binary would be around $0.24\,a_{\rm b}=5.7\,\rm au$
\citep{Paczynski1977}. However, a disc with pressure may extend to larger radii.  The larger the eccentricity of the binary, the smaller the truncation radius for the circumstellar disc
\citep{Artymowicz1994,Miranda2015}. \cite{Muller2012} find that the tidal truncation radius for a disc around $\alpha$ Centauri~B is  about $4\,\rm au$.  We consider the effect of varying the initial disc truncation radius in Section~\ref{section:rout}.

The viscosity of the disc is given by
\begin{equation} 
    \nu=\alpha c_{\rm s} H = \alpha \left(\frac{H}{R}\right)^2R^2 \Omega,
    \label{viscosity}
\end{equation}
where $\alpha$ is the dimensionless \cite{SS1973} viscosity parameter. In order to model the viscosity, we use the SPH artificial viscosity formalism \citep{Monaghan1992} that is controlled by two parameters, $\alpha_{\rm AV}$ and $\beta_{\rm AV}$. The artificial viscosity parameter $\alpha_{\rm AV}$ mimics the behaviour of an $\alpha$-disc with viscosity given by
\begin{equation}
    \alpha=\frac{\alpha_{\rm AV}}{10}\frac{\left<h\right>}{H}
    \label{alpha}
\end{equation} 
\citep{Artymowicz1994,Murray1996,Monaghan2005,Lodato2010,Meru2012,Price2018}, where $\left<h\right>$ is the mean smoothing length of particles for each spherical radius in the disc. Provided that $\alpha_{\rm AV} \gtrsim 0.1$, the physical viscosity in SPH is well resolved \citep{Bate1995,Meru2012}.  The eighth column of Table~\ref{tab} shows the artificial viscosity parameter in all of our simulations. The $\beta_{\rm AV}$ term prevents particle interpenetration \citep[e.g.][]{Monaghan1989}, and we take $\beta_{\rm AV}=2$ in all of our simulations. 

 We choose the sound speed radial profile so that the viscosity and smoothing length are both constant with radius \citep[see][]{Lodato2007}. This is a widely used choice for modelling accretion discs in SPH \citep[see also e.g.][]{Nixonetal2013,Facchinietal2013,Franchini2020,Smallwood2020}.
We choose the disc to be locally isothermal with sound speed $c_{\rm s} \propto
R^{-3/4}$.  The sound speed is constant with vertical height above the disc midplane.  With this radial profile for the sound speed, the disc scale height scales with $H =c_{\rm s}/\Omega \propto R^{3/4}$. The smoothing length therefore scales as
\begin{equation}
    \left<h\right> \propto \rho^{-1/3}\propto \left(\frac{\Sigma}{H}\right)^{-1/3} \propto R^{3/4}
\end{equation} 
where $\rho$ is the density. With equation~(\ref{alpha}), we see the $\alpha$ viscosity parameter and $\left<h\right>/H$ are
constant over the disc \citep{Lodato2007}. Thus, the disc is uniformly resolved at different radii. If we were to choose a different radial profile for the sound speed we would not be able to have a viscosity that is constant with radius. The ninth column in Table~1 shows the initial resolution of each simulation averaged over the disc.

If the disc sound crossing time-scale is shorter than the eccentricity growth time-scale, the disc is able to undergo a global response \citep[e.g.][]{Larwoodetal1996,Martinetal2014b}.  The disc aspect ratio falls off with radius as $H/R \propto R^{-1/4}$. Thus, for the fiducial model, the disc aspect ratio at $R=4\,\rm au$ is 0.059. The disc maintains radial communication  on a time-scale given approximately by $\tau=R_{\rm out}/c_{\rm s}\approx 1/[(H/R) \Omega]$ 
\citep[e.g.][]{Papaloizou1995,Lubowetal2002}, where $H/R$ is evaluated at the disc outer radius. For typical parameters, this is $\tau\approx 2.7\, \left(\frac{H/R}{0.059}\right)^{-1}P_{\rm orb}$
The larger the disc aspect ratio, the shorter the sound crossing time
and the more likely the disc is be in good communication.

We calculate the evolution of the disc properties in time. We bin the
particles into 100 radial bins and calculate the mean properties of
the particles in each bin. The inner and outer radii of the bins 
are determined by the inner and outermost particles in the simulation that are bound to the sink. For each radial bin, we calculate the surface density, $\Sigma$,  and the eccentricity vector. Thus we can find the magnitude of the eccentricity and the eccentricity apsidal angle as a function of radius in the disc. We also calculate the global density weighted disc eccentricity as
\begin{equation}
    \left<e\right>=\frac{\int_{R_{\rm in}}^{R_{\rm out}} \Sigma R e \, dR}{\int_{R_{\rm in}}^{R_{\rm out}} \Sigma R  \, dR},
    \label{ge}
\end{equation}
where $R_{\rm in}$ and $R_{\rm out}$ are the disc inner and outer radii.  Since we suggest that planet formation must take place in the inner parts of the disc we calculate this average eccentricity also just for the inner parts of the disc and define $\bar{\left<e\right>}_{\rm inner}$, as the average disc eccentricity in $R<2.5\,\rm au$.

We also define the time averaged global disc eccentricity, $\bar{\left<e\right>}$, where we typically average  from time $t=14$ to $15\,P_{\rm orb}$. This is shown in the tenth column of Table 1. Similarly we average in time the eccentricity in $R<2.5\,\rm au$ to define $\bar{\left<e\right>}_{\rm inner}$. This is shown in the eleventh column of Table 1.

\subsection{Fiducial disc around $\alpha$ Centauri~B}

For our fiducial disc model (run1b in Table~\ref{tab}) we take $R_{\rm in}=0.2\,\rm au$, the initial outer disc radius $R_{\rm out}=3\,\rm au$,
$\alpha=0.01$ and $H/R=0.1$ at $R=0.5\,\rm au$. The left panels of
Fig.~\ref{disc} show the initial density distribution  in the $x-y$ plane in which the binary orbits (upper panel) and in the $x-z$ plane (lower panel). The disc is
initially truncated at $3\,\rm au$ but it quickly spreads outwards to
about $4\,\rm au$. The other panels show the  disc once it has reached a quasi-steady state. The middle panels show the disc when the binary is at apastron at a time of $t=14.5\,P_{\rm orb}$. At this time, the disc is nearly 
axisymmetric. The right panels show the disc when the binary is at periastron at a time of $t=15.0\,P_{\rm orb}$. Strong spiral arms are induced in
the disc  at periastron that last for less than half a binary orbital period before they
are dissipated.

 The disc surface density is evolving throughout the simulation, mainly as a result of accretion onto the central binary component, $\alpha$-Centauri B. Since there is no addition of material onto the disc a steady state disc cannot be reached. The disc properties such as surface density and eccentricity are also changing on the orbital period of the binary as a result of the binary eccentricity. The disc reaches a quasi-steady state where the disc properties change in the same way each orbital period and longer period oscillations have damped down.  

 The middle and right top panels of Fig.~\ref{disc} show the formation of an inner elliptical hole in the
  gas density distribution. This is typical of isothermal 
  simulations and occurs because the particles are on eccentric orbits \citep[see also][]{Marzari2009b}. If the pericentre of a particle orbit is smaller than the sink accretion radius then the particle is accreted. In fully radiative models this hole does not appear \citep[see][]{Marzari2012,Muller2012}. 

\begin{figure*}
\begin{center}
\includegraphics[width=0.48\textwidth]{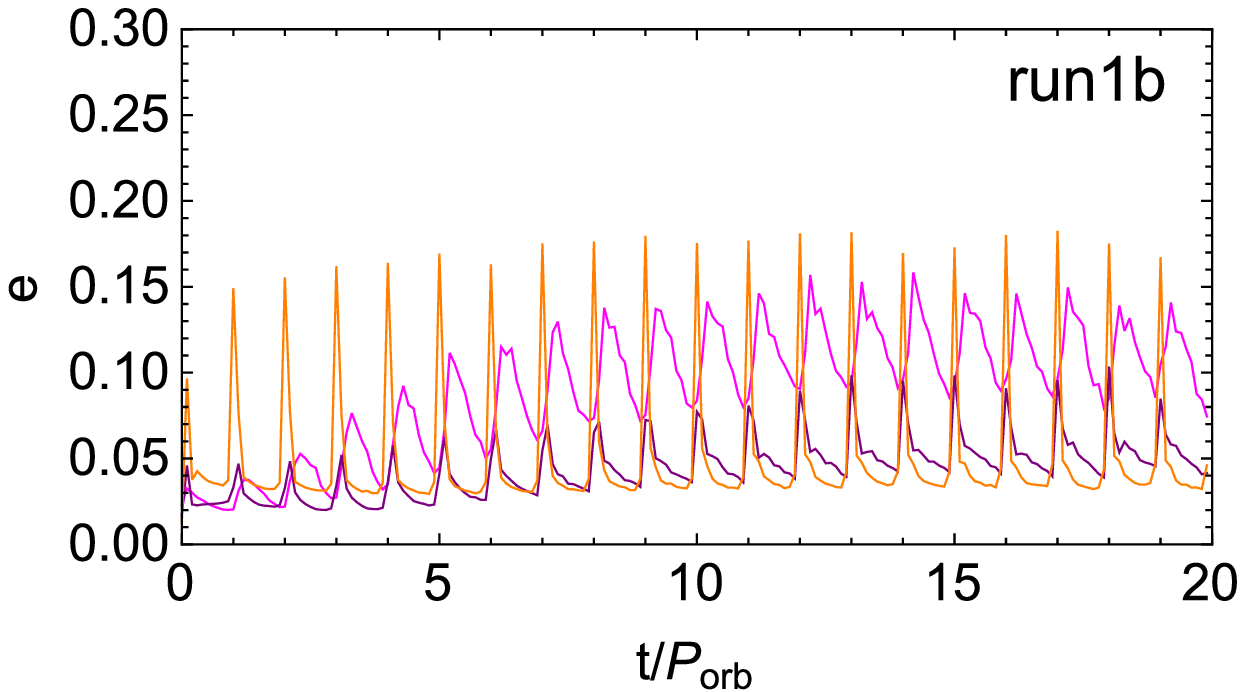}
\includegraphics[width=0.48\textwidth]{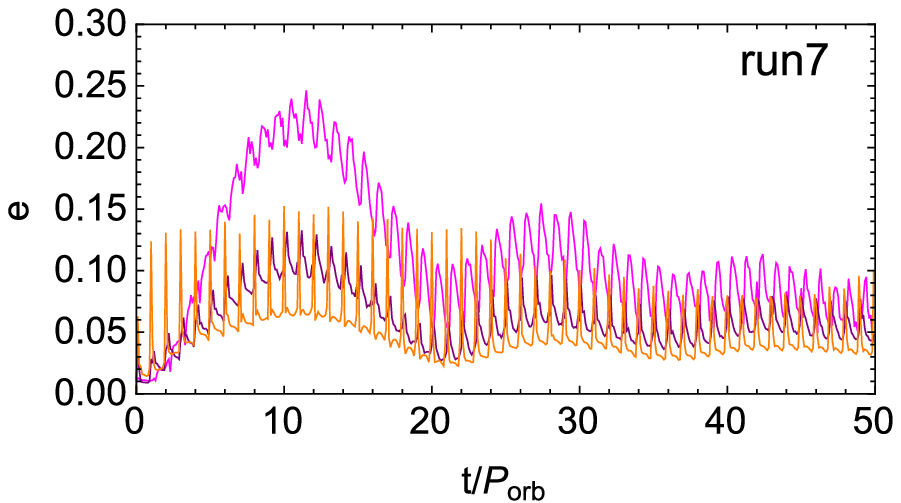}
\includegraphics[width=0.48\textwidth]{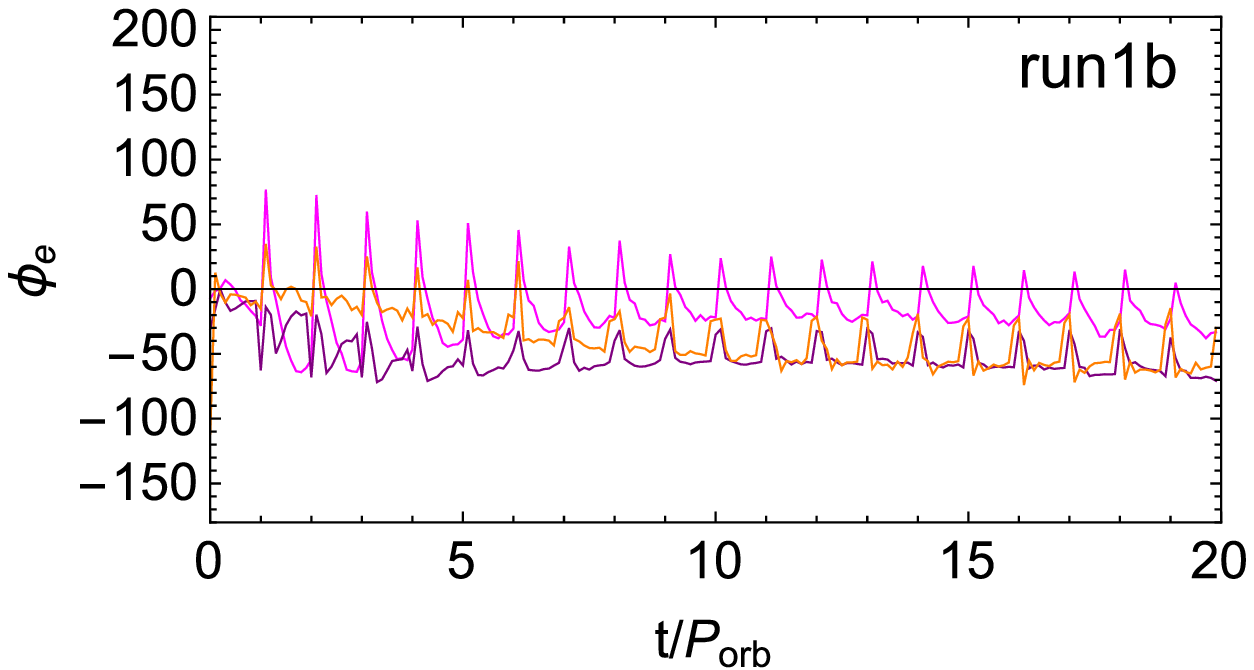}
\includegraphics[width=0.48\textwidth]{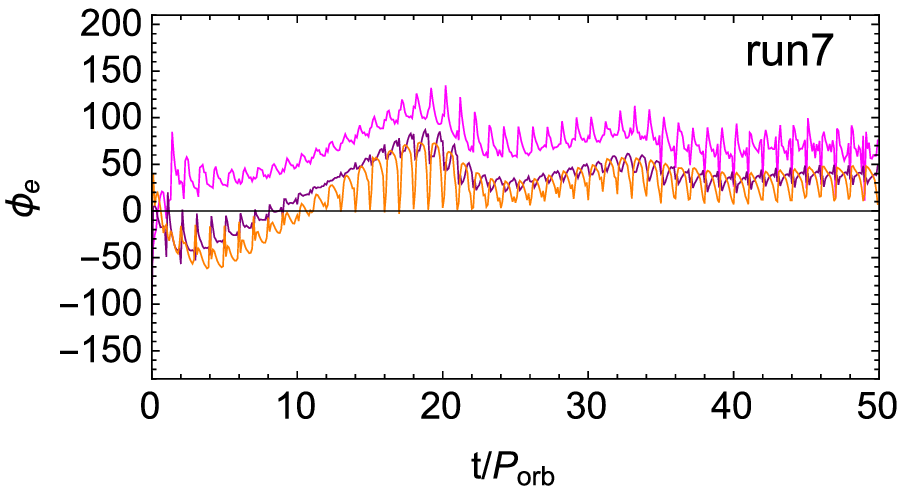}
\end{center}
\caption{Eccentricity (upper panels) and eccentricity apsidal angle (equation~\ref{phie}, lower panels) evolution of the disc at radius $R=1\,\rm au$ (magenta lines) and $R=2\,\rm au$ (purple lines) and $R=3\,\rm au$ (orange lines) for the fiducial disc model (left, run1b) and the model with  a smaller disc aspect ratio $H/R=0.05$ at $R=0.5\,\rm au$ (right, run7).}
\label{radius}
\end{figure*}

\begin{figure*}
\begin{center}
\includegraphics[width=0.48\textwidth]{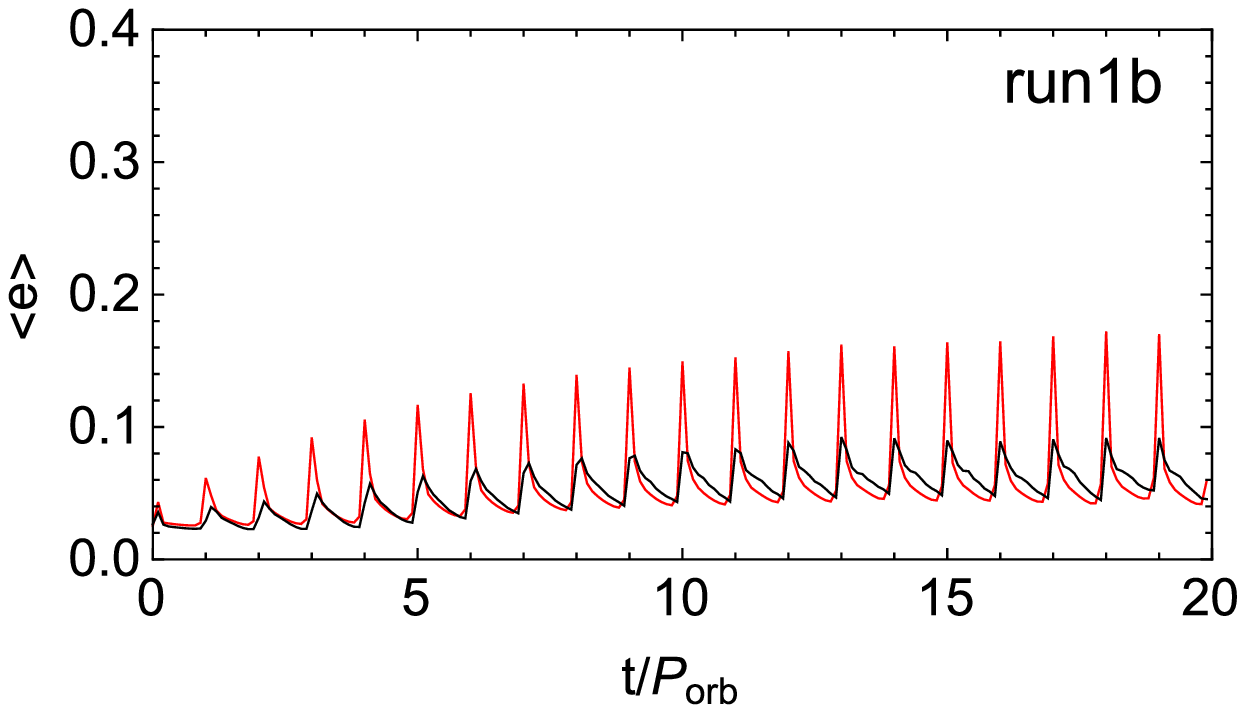}
\includegraphics[width=0.48\textwidth]{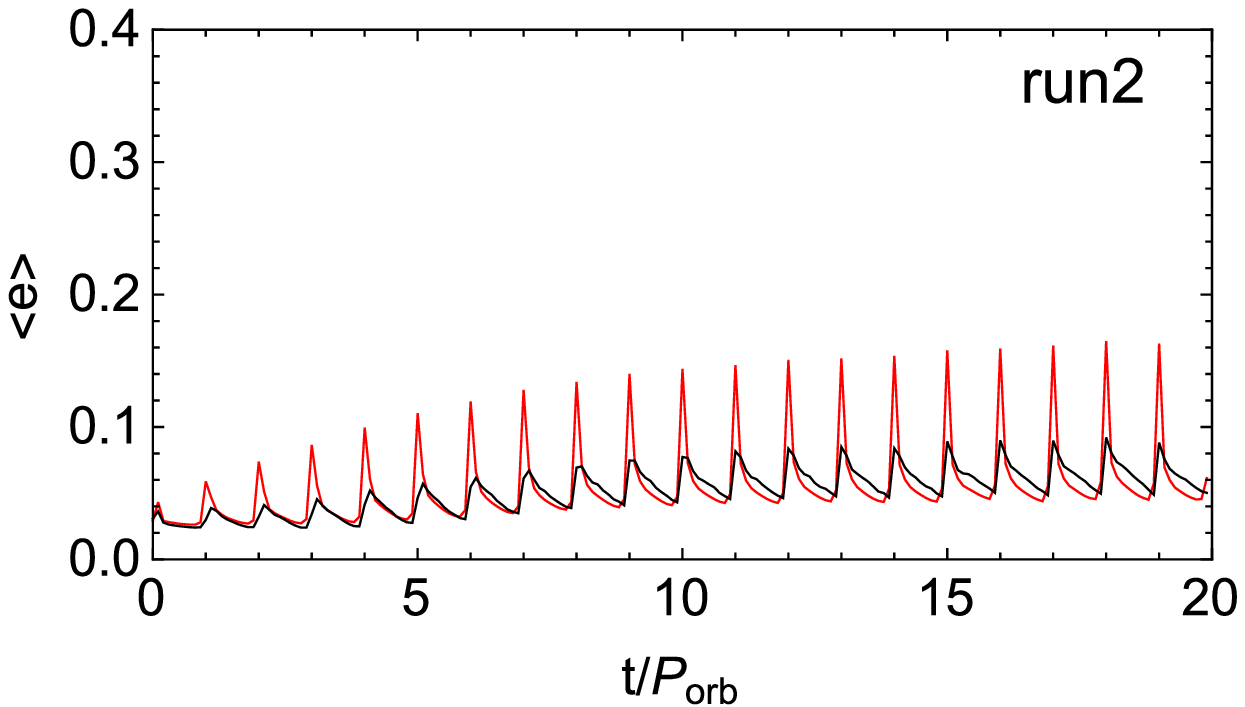}
\includegraphics[width=0.48\textwidth]{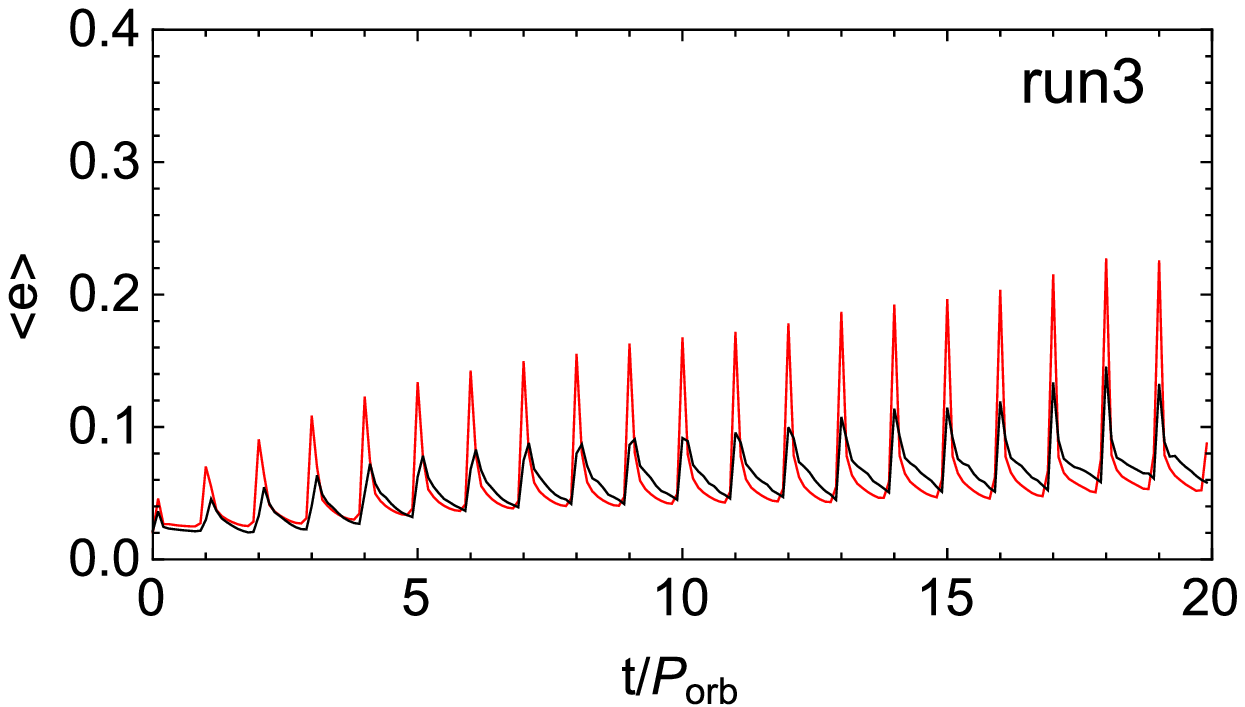}
\includegraphics[width=0.48\textwidth]{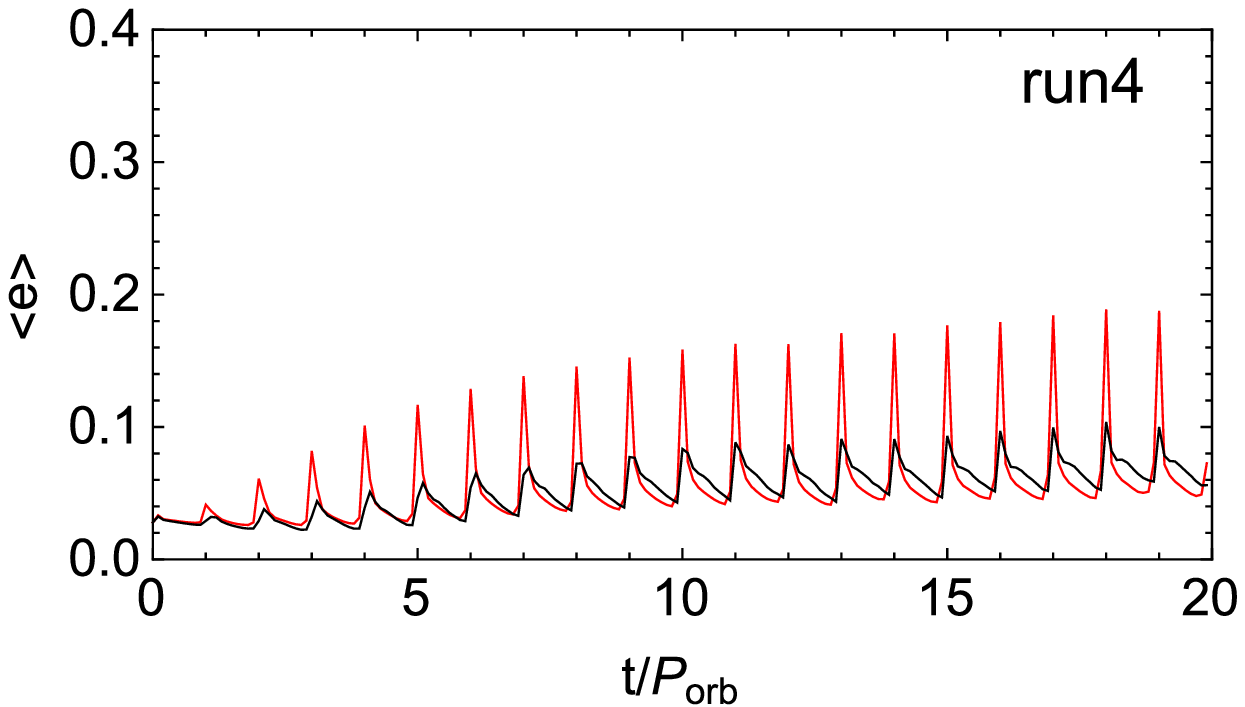}
\includegraphics[width=0.48\textwidth]{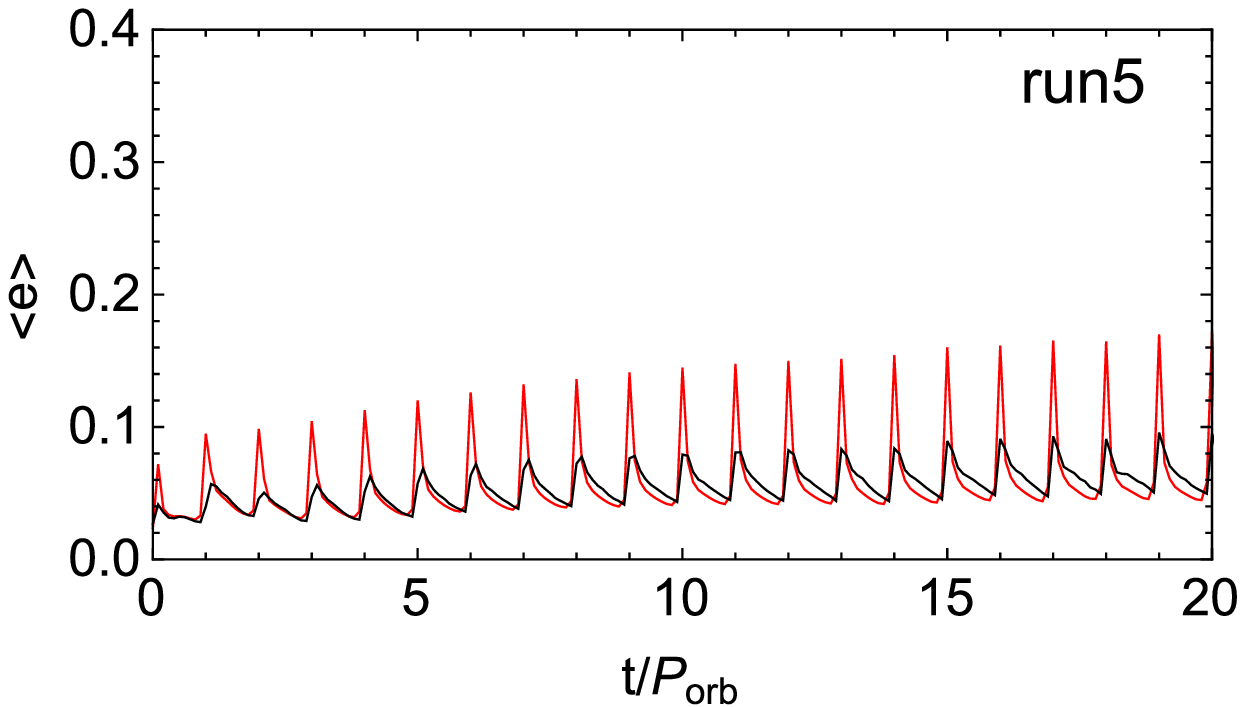}
\includegraphics[width=0.48\textwidth]{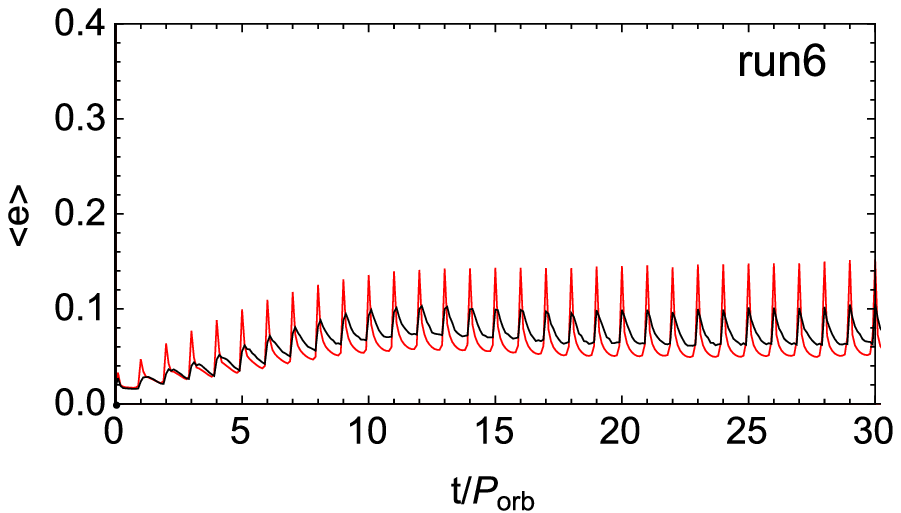}
\includegraphics[width=0.48\textwidth]{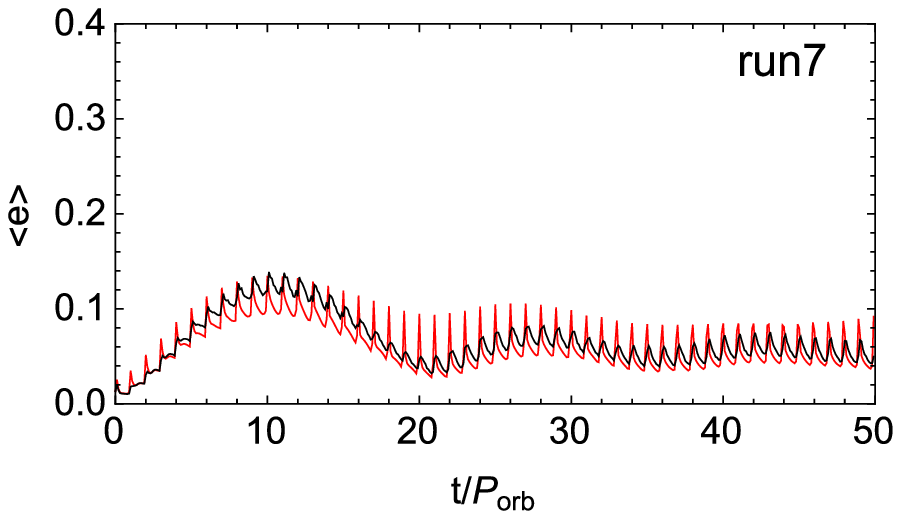}
\includegraphics[width=0.48\textwidth]{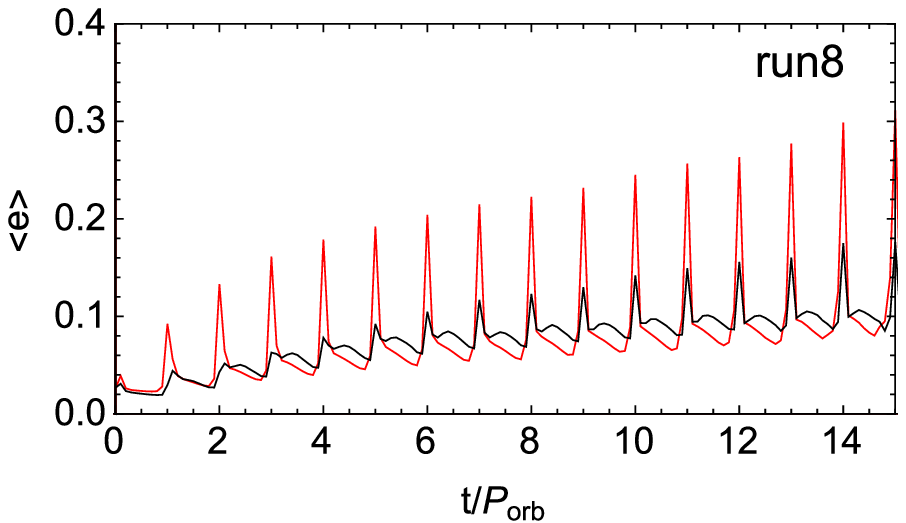}
\end{center}
\caption{Surface density weighted average eccentricity evolution of the disc. The red lines show the global disc eccentricity, $\left<e\right>$ (equation~(\ref{ge})), and the black lines show the eccentricity of the disc in $R<2.5\,\rm au$, $\left<e\right>_{\rm inner}$. 
Top left: Fiducial disc model (run1b). 
Top right: simulation with accretion radius of central star  $R_{\rm acc}=0.1\,\rm au$ (run2). 
Second row left: simulation with accretion radius of central star $R_{\rm acc}=0.5\,\rm au$ (run3). 
Second row right: simulation with initial disc outer truncation radius  $R_{\rm out}=2\,\rm au$ (run4). 
Third row left: simulation with initial disc outer truncation radius $R_{\rm out}=4\,\rm au$ (run5). 
Fourth row right: simulation with $H/R(0.5\,\rm au)=0.075$ (run6). 
Bottom row left: simulation with $H/R(0.5\,\rm au)=0.05$ (run7). 
Bottom right:  simulation with $\alpha=0.05$ (run8).
}
\label{ecc}
\end{figure*}

\begin{figure*}
\begin{center}
\includegraphics[width=0.48\textwidth]{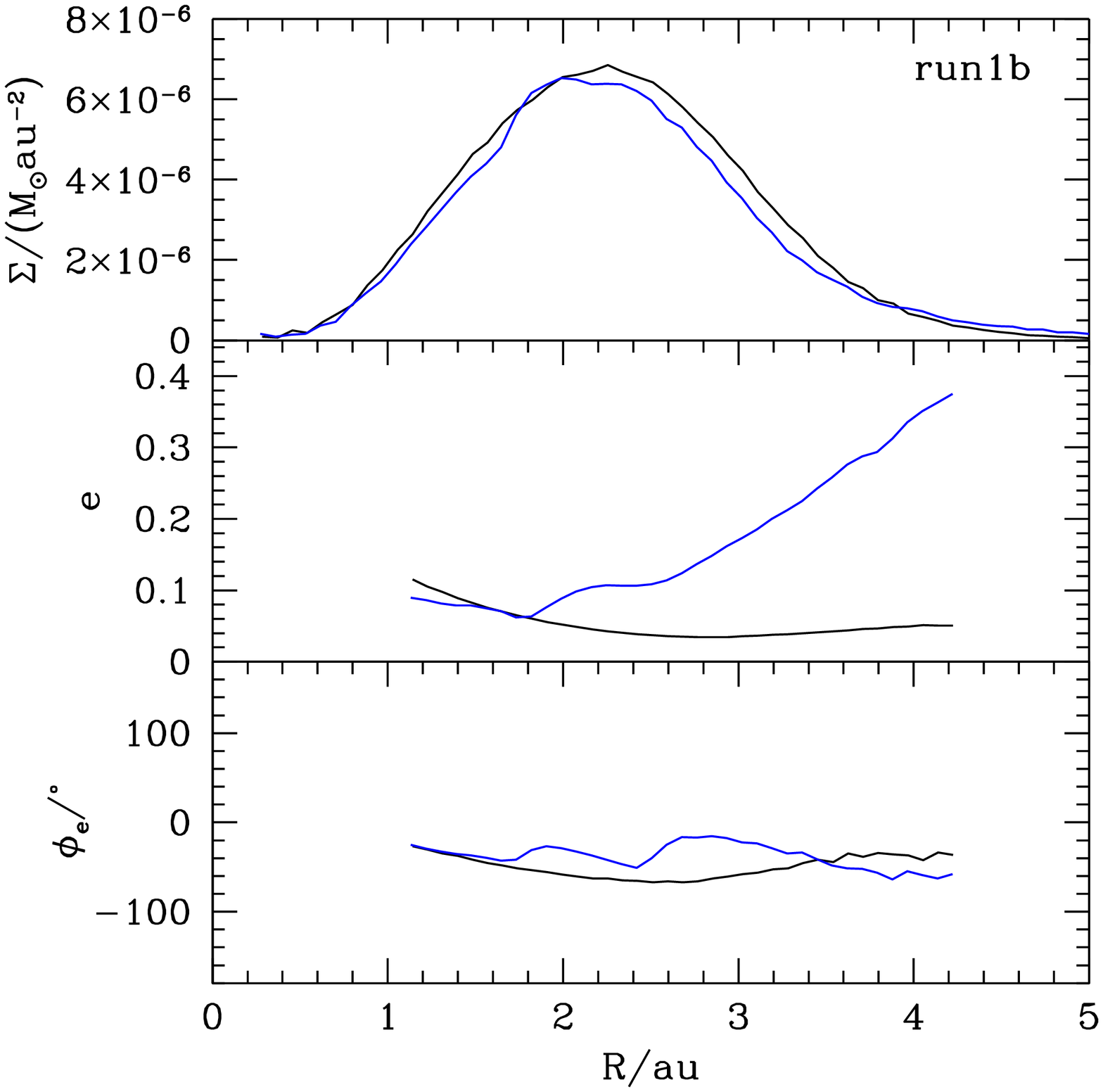}
\includegraphics[width=0.48\textwidth]{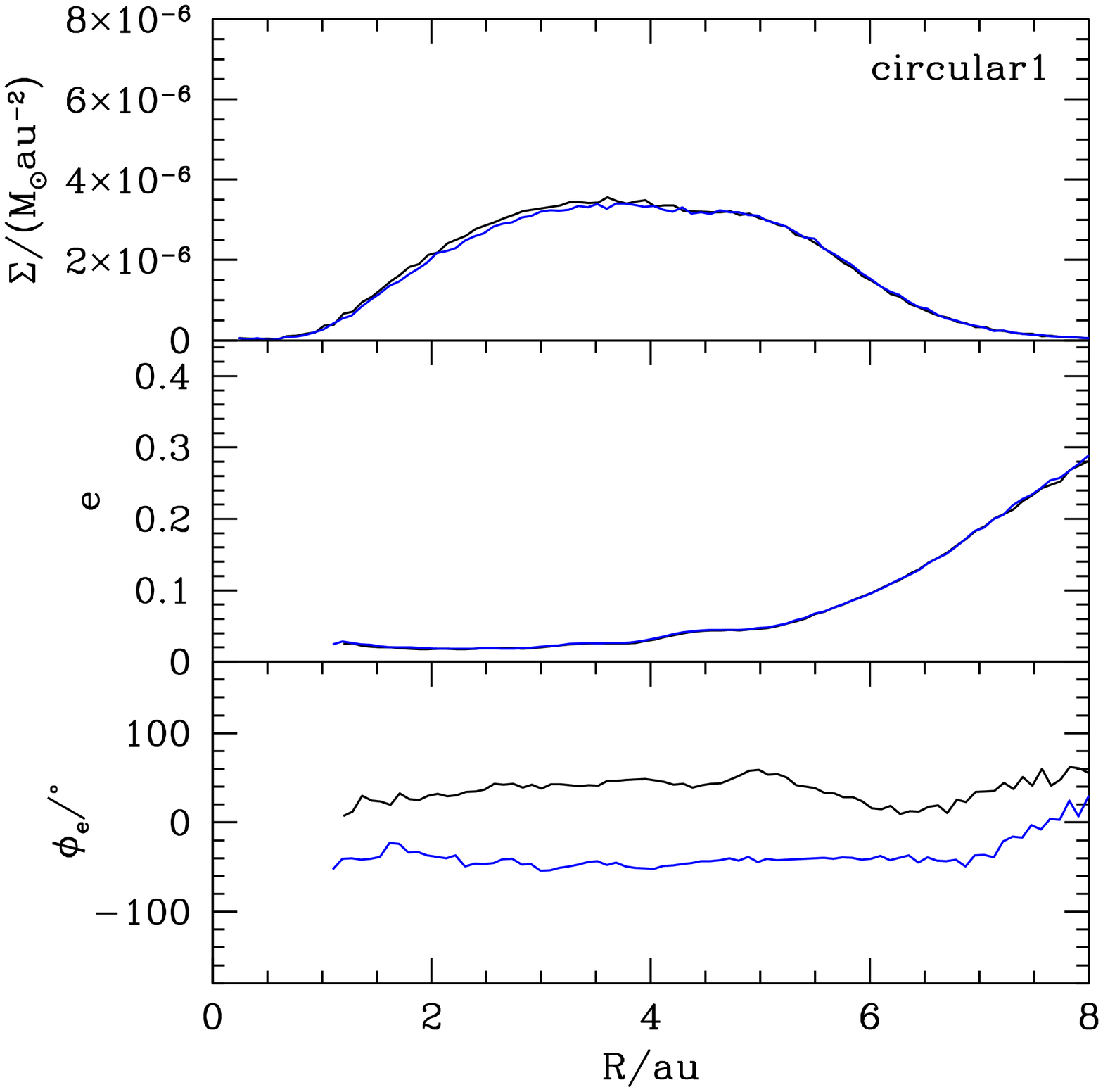}
\end{center}
\caption{Surface density (upper panels), magnitude of the eccentricity (middle panels)  and eccentricity apsidal angle (lower panels) for the fiducial disc model (left, run1b), the disc model with a circular orbit binary (right, circular1).  In each panel, the times shown are $t=14.5\,P_{\rm orb}$ (black lines) and $t=15.0\,P_{\rm orb}$ (blue lines). The eccentricity and apsidal angle lines are truncated in low density regions where $\left<h\right>/H>1$. }
\label{sd}
\end{figure*}

\begin{figure}
\begin{center}
\includegraphics[width=0.48\textwidth]{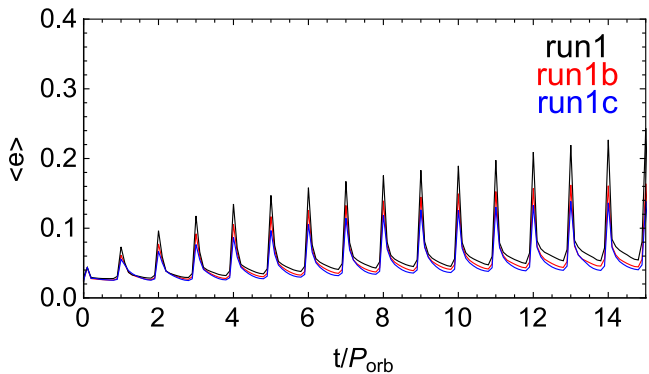}
\end{center}
\caption{Resolution study showing the globally density averaged eccentricity of the disc as a function of time for simulations with initially $1\times 10^5$ particles (black, run1), $3\times 10^5$ particles (red, run1b) and $1\times 10^6$ particles (blue, run1c). }
\label{resolution}
\end{figure}

\begin{figure}
\begin{center}
\includegraphics[width=0.48\textwidth]{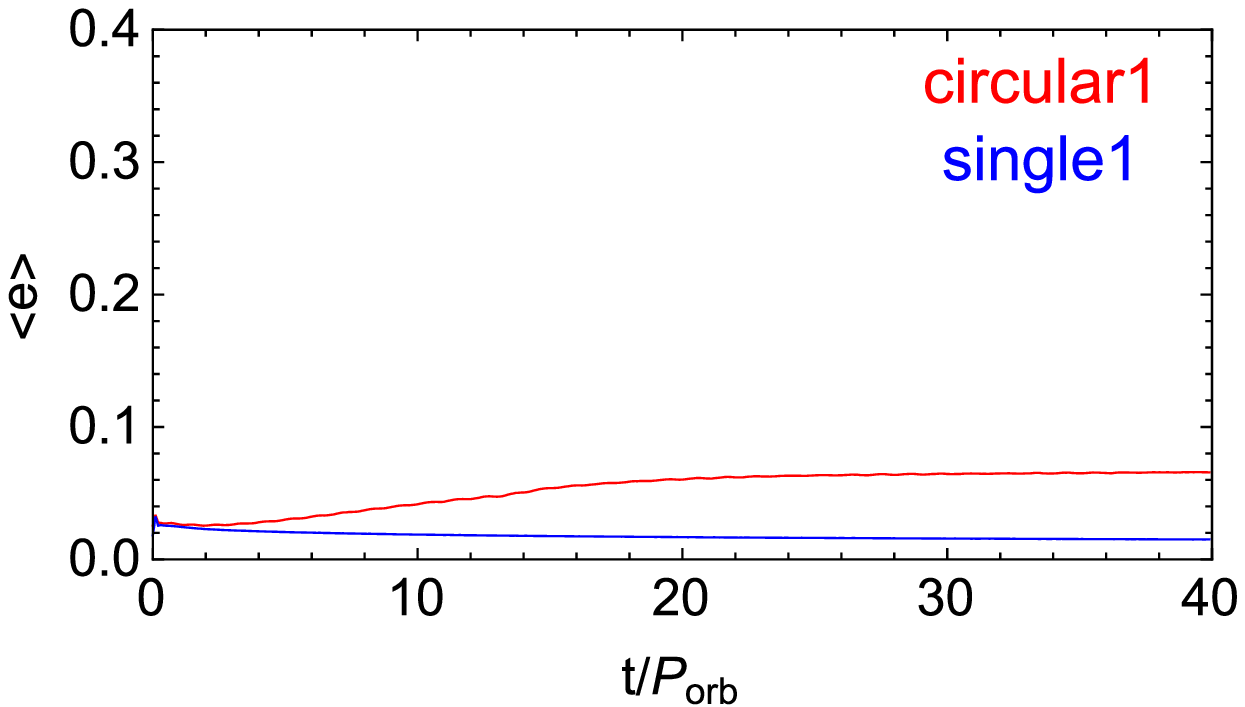}
\end{center}
\caption{Surface density weighted average global disc eccentricity  $\left<e\right>$ (equation~(\ref{ge})) evolution for the simulation without a binary companion (blue, single1) and the simulation with a circular binary companion (red, circular1).
}
\label{circsing}
\end{figure}

The top left panel of Fig.~\ref{radius} shows the time evolution of
the magnitude of the eccentricity of the disc at three different disc radii.  The disc reaches a quasi-steady state on a timescale of about 10 binary orbits.  This is in agreement with \cite{Muller2012} who found that the disc reached a quasi-steady state on a timescale of about $15\,P_{\rm orb}$.  Initially
the disc is circular at all radii. The eccentricity of the disc oscillates on the time-scale
of the orbital period of the binary at all radii reaching a local maximum when the companion is close to periastron. 
The strongest eccentricity growth is in the
outer parts of the disc (the orange lines show $R=3\,\rm au$) when the perturbing companion star is at periastron,  although it decays in a small fraction of a binary orbit. The time averaged eccentricity in the inner parts of the disc is larger ($R=1\,\rm au$ is shown in magenta and $R=2\,\rm au$ is shown in purple).
The top left panel of Fig.~\ref{ecc} shows the global density--averaged disc eccentricity calculated with equation~(\ref{ge}). The disc reaches a time averaged eccentricity of $\bar{\left< e\right>}=0.065$ at $t=14\,P_{\rm orb}$ and the inner parts have the same average eccentricity.

The bottom left panel of Fig.~\ref{radius} shows the time evolution of the disc eccentricity apsidal angle (equation~(\ref{phie})) of the disc at three different disc radii. 
A quasi-steady state is reached that oscillates only on the binary orbital period. Averaged over a binary orbital period, there is no precession in the quasi-steady state,  in agreement with \cite{Muller2012}.  The particles in the outer parts of the disc ($2-3\,\rm au$) have their periastrons aligned with one another.

 The top left panel of Fig.~\ref{sd} shows as a function of radius the disc surface density,  magnitude of the disc eccentricity  and the eccentricity apsidal angle of the quasi-steady disc at apastron (black lines, $t=14.5\,P_{\rm orb}$) 
and at periastron (blue lines, $t=15.0\,P_{\rm orb}$).  There is a peak in the surface density at a radius of about $2\,\rm au$. Interior to this radius the surface density drops because particles are accreted onto the sink particle and the disc is in a quasi-steady state \citep[e.g.][]{Pringle1981}. The density falls off outside of 2 au at a rate that is close to $R^{-3/2}$ since this is the steady state profile for the sound speed profile we have chosen. The disc is tidally truncated around $4\,\rm au$. 
At periastron there is 
strong eccentricity growth in the outer parts of the disc. The
eccentricity there increases with radius from the central star. By
apastron the eccentricity in the outer parts has been damped while the
eccentricity in the inner parts increases. There is significant eccentricity growth
in $R\lesssim 1\,\rm au$ that leads to shocks and accretion onto the
central star as there is little density there. Differential precession caused by the radial pressure gradient in the inner parts of the disc leads to a strong twisting of the disc \citep[see][]{Ogilvie2001,Ogilvie2014}. Note that high eccentricity in the inner parts of the disc has previously been seen in grid code simulations \citep{Kley2008}.

The resolution of the simulation decreases over time as particles are either accreted onto a component of the binary, form circumbinary material, or are ejected from the system \citep[e.g.][]{Franchini2019}.  At the end of the fiducial simulation, just over $10\%$ of the particles remain in the primary disc and less than a percent orbit the secondary or the binary. In total, about $85\%$ of the particles have been accreted onto the primary star during the simulation, $4\%$ have been accreted onto the secondary star and a very small fraction have been ejected from the system.  Fig.~\ref{resolution} shows a resolution study for the global eccentricity for the fiducial disc parameters with three different resolutions, $N=1\times 10^5$ particles (run1), $N=3\times 10^5$ particles (run1b) and $N=1\times 10^6$ particles (run1c). There is a slight decrease in eccentricity growth with resolution that may be attributed to the lower viscosity at higher resolution.  The global time averaged disc eccentricities shown in Table~1 do not change significantly between the two higher resolutions considered.  Thus, in order to run a large number of simulations we choose to run our simulations in the rest of this work with $3\times 10^5$ particles.

\subsection{Disc around a single star}

For comparison, we now consider the evolution of the fiducial disc model but \emph{without} the effects of the binary companion. The blue line in  Fig.~\ref{circsing} shows the eccentricity evolution in a disc around a single star with the same properties as $\alpha$~Centauri~B. As expected there is no eccentricity growth in this disc and the time averaged eccentricity at $t=14\,P_{\rm orb}$ to $t=15\,P_{\rm orb}$ is $\bar{\left< e\right>}=0.018$. We do note however that the disc spreads out to much larger radii than it does with the binary companion since there is no binary companion to truncate the disc.

\subsection{Circular orbit binary}

To remove the effect of the binary eccentricity, we also consider the evolution of the fiducial disc model with a circular orbit binary with the same binary semi-major axis as $\alpha$ Cen AB. The global eccentricity of the disc is shown in the red line in Fig.~\ref{circsing}. There is eccentricity growth in the disc, even for a circular orbit binary. This has been seen previously  \citep[e.g.][]{Muller2012}. However, there is no eccentricity oscillation on the orbital period since this is driven by the binary eccentricity. The top right panel of Fig.~\ref{sd} shows the surface density, eccentricity magnitude and eccentricity apsidal angle at a times $t=14.5\,P_{\rm orb}$ and $t=15.0\,P_{\rm orb}$. The tidal truncation of the disc in a circular orbit binary is much larger \citep[e.g.][]{Artymowicz1994, Miranda2015} and thus the disc spreads out to a radius of about $7\,\rm au$ for the circular orbit binary. There is strong eccentricity growth that increases with radius from the central star.  The larger radial extent of the disc allows for eccentricity growth through mean motion resonances \citep[e.g.][]{Lubow1991}. The global eccentricity is increasing throughout the simulation and at a time of $t=14\,P_{\rm orb}$ it reaches 0.052, which is similar to that of the fiducial model with an eccentric binary.

\subsection{Effect of the sink accretion radius}

We now consider simulations with varying accretion
radius. The top right panel of Fig.~\ref{ecc} shows a simulation for an accretion radius smaller than the
fiducial model, $R_{\rm acc}=0.1\,\rm au$ (run2) and the left panel on the second row shows a simulation for a larger accretion radius $R_{\rm acc}=0.5\,\rm au$ (run3). The disc sound speed profile is fixed between simulations.  The time averaged global disc eccentricity at the end of the simulations are $\bar{\left< e\right>}=0.063$ and $\bar{\left< e\right>}=0.071$, respectively, both very similar to the fiducial model. The simulation with the larger accretion radius leads to faster accretion of the disc and therefore poorer resolution and higher eccentricities. Comparing the evolution of the fiducial model to the the smaller accretion radius there is little difference.  Since we kept the temperature structure the same between the simulations, the accretion radius of the disc does not significantly affect the structure of the disc, just the time-scale on which the quasi-steady state disc is reached.

\subsection{Effect of the initial disc outer radius}
\label{section:rout}

The right panel of the second row of Fig.~\ref{ecc} shows a disc model with a smaller initial disc outer radius of $R_{\rm out}=2\,\rm au$ (run4). The total mass of the disc is the same as that of the fiducial disc, as is the mass of individual particles; only the initial locations and velocities of the particles differ. The evolution of the disc is very similar to the fiducial disc model. However, the disc lifetime is shorter since the particles are initially distributed much closer to the central star.

We also consider a disc with a larger initial truncation radius. The left panel of the third row of Fig.~\ref{ecc} shows a simulation with an initial outer truncation radius of $R_{\rm out}=4\,\rm au$ (run5). The spikes in the eccentricity are larger at the start of the simulation since there is more material in the outer regions. However, the quasi-steady state eccentricity of the disc is very similar to that of the simulations with smaller initial truncation radii.  The time averaged global disc eccentricity for the simulations with $R_{\rm out}=2\,\rm au$ (run4), $3\,\rm au$ (run1b) and $4\,\rm au$ (run5) are very similar, and thus the initial disc outer radius does not significantly affect the disc evolution or the final quasi-steady state disc structure.  The truncation radius of the disc in the quasi-steady state is around $4\,\rm au$, as shown in the surface density panel in Fig.~\ref{sd}. This is independent of the initial truncation radius of the disc and in agreement with the simulations of \cite{Muller2012}.

\subsection{Effect of the disc aspect ratio, $H/R$}

Fig.~\ref{ecc} shows simulations with a smaller disc aspect ratio
than the fiducial model, $H/R(R=0.5{\,\rm au})=0.075$ (run6, third row right) and $H/R(R=0.5{\,\rm au})=0.05$ (run7, bottom left).  A smaller disc aspect ratio leads to a lower disc viscosity (see equation~(\ref{viscosity})). 
For run7 in particular, since there is less viscosity in the disc (and hence eccentricity damping), the disc undergoes two types of oscillations. The first are on the binary orbital period, as seen in the fiducial disc model. The second type of eccentricity oscillations occur on a timescale of about $20 \,P_{\rm orb}$. 
These oscillations are due to the forced
eccentricity driven by the companion star, as experienced by a test
particle as shown  in Fig.~\ref{particle}.  
The
disc is able to hold itself together and undergoes global oscillations
rather than each radius oscillating on a different time-scale.  The
test particle orbits display an increasing eccentricity growth with radius
from the central star. However, the inner parts of the disc show the opposite
behaviour. The right panels of Fig.~\ref{radius} show for three different disc radii the eccentricity and eccentricity apsidal angle evolution. Each radius undergoes eccentricity oscillations on the same timescale but with varying magnitudes. The inner parts of the disc undergo larger magnitude oscillations. This is opposite to the test particles and a result of the communication across the disc.  

For large timescales, the global eccentricity oscillations damp out due to the viscosity of the disc and the disc eccentricity converges to a value that oscillates only on the orbital time-scale. The time averaged global disc eccentricity at the end of the run7 is 0.050 and this is lower than the fiducial model. Thus, a lower disc aspect ratio may lead to higher disc eccentricity initially while the global forced eccentricity oscillations operate, but it leads to a smaller disc eccentricity in the quasi-steady state disc.

\subsection{Effect of the disc viscosity $\alpha$}

The bottom right panel of Fig.~\ref{ecc} shows a simulation with higher $\alpha$ than the
fiducial model, $\alpha=0.05$ (run8).  Note that this simulation does not reach a quasi-steady state because the evolution of the disc is much faster with a higher viscosity parameter. The eccentricity growth is larger  for larger viscosity. The time averaged global disc eccentricity at a time of $14-15\,P_{\rm orb}$ is $\bar{\left< e\right>}=0.12$, higher than  the fiducial model. This is in agreement with grid code simulations that showed that a larger viscosity leads to a larger disc eccentricity \citep[e.g.][]{KPO2008,Muller2012}.  Thus, a low disc eccentricity requires a low viscosity. This is achieved through both a low viscosity $\alpha$ parameter and a low disc aspect ratio, as shown by equation~(\ref{viscosity}).

\begin{figure}
\begin{center}
\includegraphics[width=0.45\textwidth]{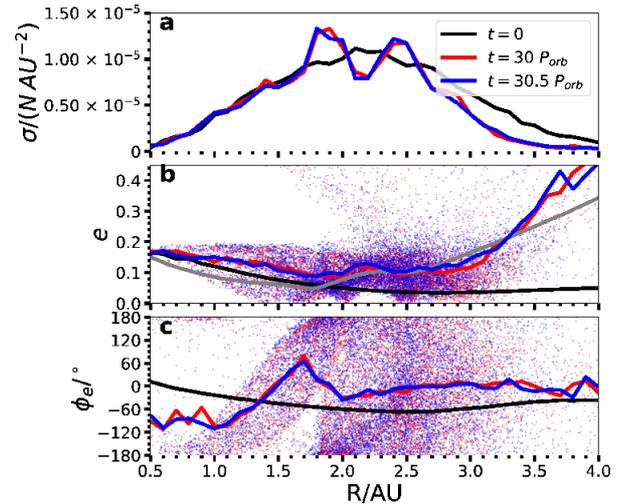}
\end{center}
\caption{Surface density (panel a), magnitude of the eccentricity (panel b)  and eccentricity apsidal angle (panel c) for a disc of test particles orbiting $\alpha$ Centauri~B. The binary is at apastron at $t=0$. The initial surface density and particle eccentricity is taken to be the steady state achieved in the gas disc through our hydrodynamical simulation \textit{run1b}. The black lines show the binned average initial values.   The  points show individual test particles at apastron at time $t=30\,P_{\rm orb}$ (red) and at periastron at time $t=30.5\,P_{\rm orb}$ (blue).  The red and blue curves show the corresponding binned and averaged values. The grey curve marks the eccentricity profile for the steady state solution from the hydrodynamical simulation Fig. \ref{sd} at periastron. 
}
\label{sd_test}
\end{figure}

\section{Particle disc simulations}
\label{discussion}

The hydrodynamical disc simulations presented in this work show that a gas disc around $\alpha$ Centauri B cannot remain azimuthally symmetric. The radial eccentricity structure of the disc (see Fig.~\ref{sd}) is somewhat different from the forced eccentricities of test particles (see Fig.~\ref{particle2}) that represent a noncollisional planetesimals disc in the absence of gas. In order to understand the orbital properties of planetestimals that form from the gas disc, we now consider a disc of test particles that interact with the binary only through gravity using the n-body GPU code, \texttt{GENGA} \citep{Grimm2014}. The goal of these simulations is to understand the evolution of a disc of planetesimals that are assumed to form with the same orbital properties of the gas disc. The planetesimals do not interact with the gas disc, only with the gravity of the binary.

We consider a particle disc of $10^4$ test particles. The particles are initially distributed so that the particle density (number of particles per unit area) has the same distribution with radius as the surface density of the gas disc in run1b at a time of $t=15\,P_{\rm orb}$ (see the left panel of Fig. \ref{sd}).  The eccentricity magnitude and eccentricity apsidal angle are also given by the gas disc elements from the hydrodynamical simulations in run1b.  
The particles all orbit initially coplanar to the binary orbit.
 The binary orbit starts from apastron and the system is evolved for 30 $P_{\rm orb}$.  For comparison to the gas disc simulations, we define the particle surface density as the number of particles per unit area, $\sigma$. We bin the particles into 60 radial bins and average the eccentricity and apsidal phase angle within each bin.

We assume that the planetesimals form in the disc with the orbital properties of the disc, but they are initially on Keplerian orbits. However, the gas disc is partially pressure supported, meaning that the gas orbits at slightly sub-Keplerian velocity. The azimuthal velocity of the gas is 
\begin{equation}
   \frac{ v_\phi^2}{R}=\frac{G M_B}{R^2} +\frac{1}{\rho}\frac{dP}{dR}
\end{equation}
\citep{Pringle1981,Armitage2015},
where $\rho$ is the gas density and $P$ is the pressure. Considering the hydrodynamic simulations presented in this work,  the disc aspect ratio scales as $H/R \propto R^{-1/4}$ and the sound speed as $c_{\rm s}\propto R^{-3/4}$. Assuming that the surface density distribution is a power law that scales as $\Sigma \propto R^{-x}$,  the density scales as $\rho \propto \Sigma/H \propto R^{-(x+3/4)}$. The pressure $P=c_{\rm s}^2\rho$ scales as $P\propto R^{-(x+9/4)}$ 
The relative difference between the disc azimuthal velocity and Keplerian velocity is
\begin{equation}
    \frac{v_{\rm \phi}-v_{\rm K}}{v_{\rm K}}\approx -\frac{1}{2}(x+9/4)\left(\frac{H}{R}\right)^2.
\end{equation}
We can find an upper limit to this by taking the  initial surface density power law with $x=3/2$. However, we note that the surface density evolves significantly from a power law distribution towards lower values of $x$. Even for the initial conditions, the difference in velocities at $R=1\,\rm au$ has a maximum of about 1\%.  This difference is important for planet formation \citep{Youdin2005,Youdin2007,Yang2014} but since we do not include the interaction between the gas disc and the planetesimals in our N-body calculations, we ignore this effect. The particles begin with Keplerian velocity with the eccentricity distribution of the gas disc.   If the particles took on the gas velocity, rather than the gas  eccentricity, the semi-major axis of the particles would initially drop as the particles would fall inwards, with closer mean orbital radius. However, this effect is small. 

We do not include the effects of gas drag on the planetesimals. Gas drag may lead to pericenter alignment of the orbits and an equilibrium distribution for the eccentricity of small planetesimals.  The gas drag force depends on the planetesimal size and so the variation in the magnitude of the eccentricity and the apsidal angle with particle size may excite large mutual impact velocities and inhibit the formation of larger bodies \citep{Marzari2000,Thebault2006,Thebault2008,Thebault2009,Paardekooper2008,Xie2009}. In an eccentric disc, the relative particle velocities are increased compared to a circular orbit disc \citep{Paardekooper2008}. \cite{Marzari2012} found that the eccentricity of planetesimals that are influenced by gas drag  are significantly higher than those driven only by the secular perturbations of the binary companion. Even when gas drag is included, the perturbations by the eccentric disc excite high planetesimal eccentricities, especially in the inner parts of the disc.

 \subsection{Test particle simulation} \label{sec:test_part}

We first consider a test particle simulation in which the particles do not interact with each other. Figure \ref{sd_test}a illustrates that the evolved azimuthally-averaged surface density profile (red and blue) is similar to the initial state (solid black line). A gas disc is able to extend farther out than the test particles because of the pressure in the gas disc that the test particles do not feel. 
Thus, there is a decrease in the density of particles in the outer regions compared to the initial distribution.
The magnitude of the eccentricity for the middle-to-outer portion of the particle disc ($\gtrsim$2 au) in Fig. \ref{sd_test}b is substantially excited (red and blue lines) relative to the initial state (solid black line) and in a similar fashion as the excitation of the gas disc due to the binary companion at periastron (Fig. \ref{sd_test}b grey line).  Beyond $3 \,\rm au$, the particles experience a large eccentricity excitation and are removed over subsequent encounters with the stellar companion.

The radially-averaged eccentricity apsidal angle of the test particles is depicted in Fig. \ref{sd_test}c for both the initial (black) and the evolved (red and blue) state.  Each radial bin consists of a broad range of eccentricity apsidal angles, where the expectation, if uniformly distributed, is $\phi_e \sim 0^\circ$.  The initial eccentricity of the particles is always larger than the forced eccentricity (see Fig.~\ref{particle2}) and therefore all particles are circulating, meaning that they precess through the full range of values for $\phi_{\rm e}$. 
The eccentricity apsidal angles for the particles become randomly distributed in the particle disc since each precesses on a timescale depending on the distance from the binary companion.   The outer portion of the particle disc ($\gtrsim 2.0$ au) averages to approximately zero in Fig. \ref{sd_test}c since the precession is fastest there. Over a longer period of time, the whole disc will become randomly distributed. This is in contrast to the gas disc which has a smoothly varying eccentricity apsidal angle throughout the disc because the disc is radially connected through pressure.

From Fig.~\ref{particle2}, we expect the forced eccentricity to represent up to about half of the eccentricity magnitude (blue and red lines) shown in Fig. \ref{sd_test}b until $\sim$3 au, where the remainder lies in a free eccentricity component. The average eccentricity of the particles in $R<2.5\,\rm au$ increases from 0.088 initially up to 0.14 after a time of 30 binary orbits. There is little difference in the average eccentricity at binary periastron and apastron. This suggests that  the gas disc actually helps to suppress the magnitude of the eccentricity of particles within it if they are well coupled to the gas. Furthermore, the gas disc may allow coupled particles to have aligned periastrons. Planet formation in $R\gtrsim 2.5\,\rm au$ may be difficult due to the significant eccentricity growth in the outer parts of both the gas and particle discs. 

\subsection{Effect of mutual planetesimal interactions}

\begin{figure*}
\begin{center}
\includegraphics[width=0.8\textwidth]{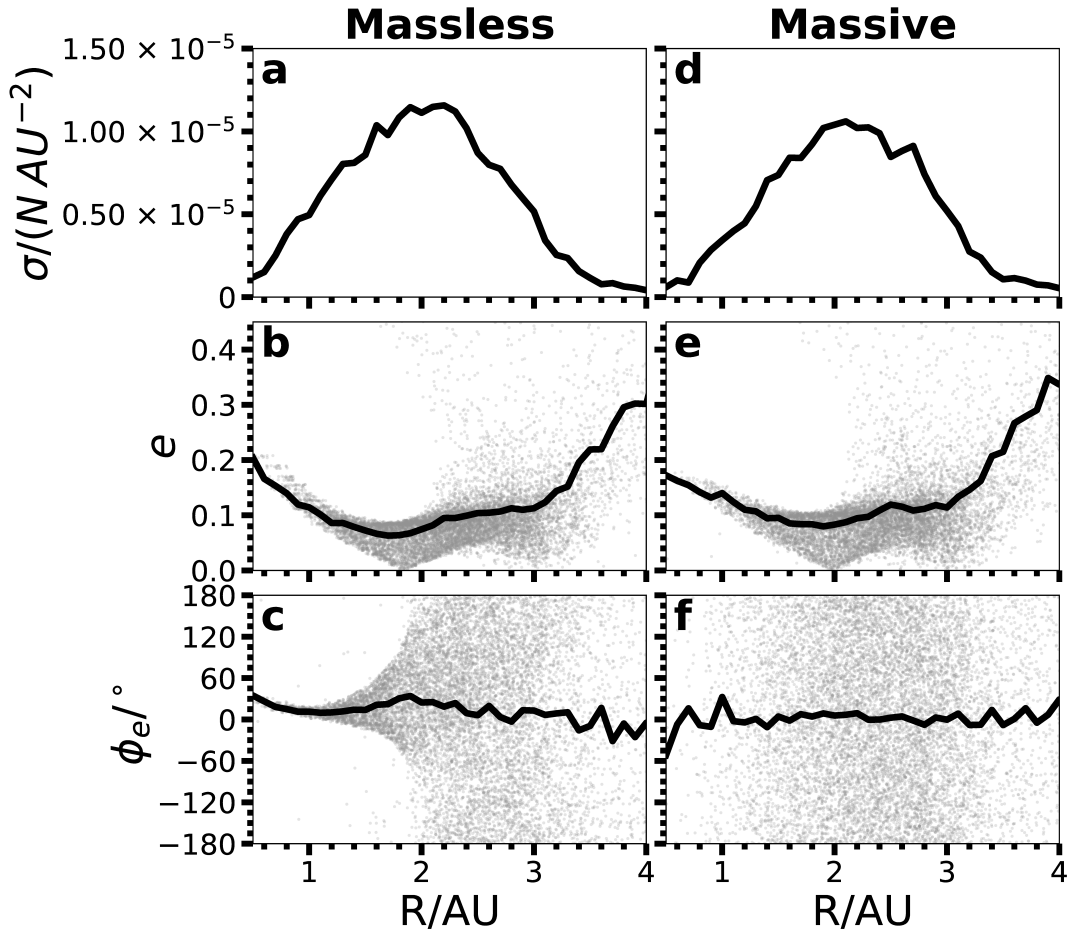}
\end{center}
\caption{Particle simulations showing the effect of massive particles.  Upper panels show the particle disc surface density, the middle panels show the magnitude of the eccentricity and the lower panels show the eccentricity apsidal angle. The initial surface density and particle eccentricity is taken to be the steady state achieved in the gas disc through our hydrodynamical simulation \textit{run1b}. The black lines show averaged values and the grey points show each individual particle.  Left panels (a--c): The same simulation as shown in Fig.~\ref{sd_test} except the time shown is $t=4.\,P_{\rm orb}$. These panels show a disc of massless test particles. Right panels (d--f): A disc of massive (0.0006 M$_\oplus$) interacting particles at time $t=4.\,P_{\rm orb}$. }
\label{collisions}
\end{figure*}

We now consider the effect of mutual planetesimal interactions on the evolution of the planetesimal orbits. We run a simulation with the same initial conditions as in Section \ref{sec:test_part}, except we replace the test particles with massive (0.0006 M$_\oplus$) particles resulting in a total disk mass of $\sim$6 M$_\oplus$.  These particles interact  gravitationally with each other, which allows for additional effects (scattering or dynamical friction) that might alter the alignment of the particles.  Simulations of this type are computationally expensive (even with GPU parallelization), so we simulate only for a few binary orbits (4 $P_{\rm orb}$ = 320 yr).  This timescale is sufficient to capture several close approaches of the stellar companion, which induces apsidal precession, and allows for the possibility of  self-stirring of the particles due to their own mutual interactions.

 Figure ~\ref{collisions} shows a comparison between our simulations with and without particle interactions at the same epoch (4 $P_{\rm orb}$). Similar to Fig. \ref{sd_test}, we show the surface density, eccentricity, and apsidal alignment profiles, where the gray points mark the instantaneous values for each particle.  The magnitude of the eccentricity radial profile (Fig. \ref{collisions}b and \ref{collisions}e) is similar between the two simulations, where Fig. \ref{collisions}e shows a slightly lower eccentricity within the inner disc and a minimum eccentricity occurring further out (2 au). The main effect of the mutual interactions is seen in the eccentricity apsidal angle in the lower panels (Fig. \ref{collisions}c and \ref{collisions}f). The mean pericenter is close to zero because the particles scatter off of each other.  As a result, the pericenters have become uniformly distributed in the range $-180^\circ$ to $180^\circ$. Thus, the effect of the particle interactions is that the planetesimal disc has reached a quasi-steady state on a shorter timescale.

\section{Discussion and Conclusions}
\label{conc}

An initially circular protoplanetary disc around one component of an eccentric binary may
display two types of eccentricity oscillations. First, the eccentricity oscillates on the orbital period of the binary. Second, global forced eccentricity oscillations occur on a longer time-scale if the disc aspect ratio is sufficiently small. For a disc around $\alpha$ Centauri~B, the oscillations are on a time-scale of around $20\,P_{\rm orb}$. The oscillations
damp through viscous dissipation until a quasi-steady state is reached. The eccentricity of the
disc is then excited in the outer parts of the disc with each orbital
period. In the outer parts of the disc the eccentricity increases with
distance from the central star. However, close to the star, the
eccentricity decreases with increasing radius. 

Large eccentricities in a disc of planetesimals may  make planet formation difficult due to increased relative velocities between planetesimals and thus destructive collisions. We have shown that the quasi-steady state global gas disc eccentricity magnitude is lower for smaller viscosity $\alpha$ parameter and smaller disc aspect ratio, $H/R$. Due to computational restrictions, we are unable to simulate a well resolved  disc with smaller $\alpha$ or smaller $H/R$ than those presented. However, we suggest that a small viscosity $\alpha$ parameter and small $H/R$ are required for planet formation around $\alpha$ Centauri B. 

 In the gas disc, the periastons of the gas particle orbits reach a quasi-steady state in which they are nearly aligned to each other and exhibit no precession. If the presence of the gas disc causes periastron alignment of the particle orbits then the impact velocities may be small even in an eccentric disc \citep[e.g.][]{Thebault2004}.  
Assuming that planetesimals form with similar properties  to the gas disc, we found that after the gas disc is dispersed, or the particles become large enough to decouple from the gas disc, the average eccentricity magnitude of the particles increases and the periastrons become misaligned to each other.  Thus, the presence of  gas may make planet formation around $\alpha$ Centauri B easier by allowing for planetesimal collisions at lower relative velocities.

In this work we have assumed that the disc is coplanar to the orbit of
the binary.  The relatively small binary separation may lead to
alignment of the disc to the binary orbital plane during the disc
lifetime \citep[e.g.][]{Bateetal2000}. However, observations of discs
in binary star systems suggest that large misalignments between the
angular momentum of the disc and the angular momentum of the binary
may be common
\citep[e.g.][]{stapelfeldt1998,Jensen2014,Williams2014}. The tidal
torque on a disc that is inclined to the binary orbital plane is
weakened and thus the disc can viscously spread farther out
\citep{Lubowetal2015}.  Forced eccentricity
oscillations are smaller in a misaligned disc.  For high disc misalignments, the disc
may be unstable to Kozai--Lidov \citep{Kozai1962,Lidov1962}
oscillations where the disc eccentricity and inclination are exchanged
\citep{Martinetal2014,Fu2015a,Fu2015b}, and this will influence planet formation
processes in the disc.  Note that planet formation through fragmentation may be enhanced in a disc undergoing KL oscillations \citep{Fu2017}.
The alignment time-scale of a misaligned circumprimary disc around $\alpha$ Centauri is short compared to the lifetime of the disc. However, circumstellar discs may be fed material through a circumbinary disc \citep[e.g.,][]{Nixonetal2011a,Bate2018,Alves2019}. If the circumbinary disc is misaligned to the binary orbit, then the circumstellar disc may remain misaligned.

\section*{Acknowledgments} 

We thank an anonymous referee for useful comments and a thorough review. We thank Daniel Price for providing the {\sc phantom} code for SPH
simulations and acknowledge the use of SPLASH \citep{Price2007} for
the rendering of the figures. RGM acknowledges support from NASA through grant NNX17AB96G. Computer support was provided by UNLV's
National Supercomputing Center.  This research was supported in part through research cyberinfrastructure resources and services provided by the Partnership for an Advanced Computing Environment (PACE) at the Georgia Institute of Technology.

\bibliographystyle{mnras}

\bsp
\label{lastpage}
\end{document}